%% file: main.tex
\documentclass{article}

\input{./preamble}

\input{./paper_preamble}

\renewcommand{\vec}{\bm}
\newcommand{\m}{\mathrm{m}}
\newcommand{\A}{\mathrm{A}}
\newcommand{\B}{\mathrm{B}}
\newcommand{\rot}[1]{\mathfrak R (\theta_{#1})}
\newcommand{\Hchi}{H_\mathrm{chiral}}
\newcommand{\Hbm}{H_\mathrm{BM}}
\newcommand{\Hrel}{H_\mathrm{relax}}

\newcommand{\HbmVV}{H_\mathrm{BM,valleyful}}
\newcommand{\HrelVV}{H_\mathrm{relax,valleyful}}

\newcommand{\eps}{\varepsilon}
\newcommand{\dee}{\ensuremath{\textrm{ d}}}
\usepackage{authblk}

\graphicspath{{images/}}
\usepackage[labelformat=simple]{subcaption}	
\usepackage[margin=1in]{geometry}
\usepackage{lscape}

\title{Interacting Twisted Bilayer Graphene with Systematic Modeling of Structural Relaxation}

\author[1]{Tianyu Kong}
\author[2]{Alexander B. Watson}
\author[3]{Mitchell Luskin}
\author[4]{Kevin D.  Stubbs}

\affil[1,2,3]{School of Mathematics, University of Minnesota, Minneapolis, MN, 55455, USA\\kong0226@umn.edu,\ abwatson@umn.edu,\ luskin@umn.edu}
\affil[4]{Department of Mathematics, University of California, Berkeley, CA 94720, USA\\kstubbs@berkeley.edu }

\date{\today}
\begin{document}
\maketitle

\begin{abstract}
  Twisted bilayer graphene (TBG) has drawn significant interest due to recent experiments which show that TBG can exhibit strongly correlated behavior such as the superconducting and correlated insulator phases.
  Much of the theoretical work on TBG has been based on analysis of the Bistritzer-MacDonald model which includes a phenomenological parameter to account for lattice relaxation.
  In this work, we use a newly developed continuum model which systematically accounts for the effects of structural relaxation.
  In particular, we model structural relaxation by coupling linear elasticity to a stacking energy that penalizes disregistry.
  We compare the impact of the two relaxation models on the corresponding many-body model by defining an interacting model projected to the flat bands.
  We perform tests at charge neutrality at both the Hartree-Fock and Coupled Cluster Singles and Doubles (CCSD) level of theory and find the systematic relaxation model gives quantitative differences from the simplified relaxation model.
  
\end{abstract}
\section{Introduction}


Recent experiments on correlated electrons in twisted bilayer graphene (TBG) have shown a rich many-body electronic phase diagram with many competing candidate many-body ground states \cite{NuckollsYazdani2024,NuckollsLeeOhEtAl2023,BultinckKhalafLiuEtAl2020,OhNuckollsWongEtAl2021}. This has motivated intense activity to develop reduced-order models capable of capturing this behavior\cite{SongBernevig2022,CalugaruBorovkovLauEtAl2023,CancesGarrigueGontier2023,Kong_Liu_Luskin_Watson_2024,Quinn_Kong_Luskin_Watson_2025,Carr_Fang_Zhu_Kaxiras_2019,Fang_Carr_Zhu_Massatt_Kaxiras_2019,Watson_Kong_MacDonald_Luskin_2023}. These reduced-order models generally take the form of a moir\'e-scale PDE modeling TBG's single-particle electronic properties, such as the well-known Bistritzer-MacDonald model \cite{Bistritzer_MacDonald_2011}, coupled to a Coulomb electron-electron term, projected down to the single-particle model's flat moir\'e-Bloch bands \cite{BultinckKhalafLiuEtAl2020,ChatterjeeBultinckZaletel2020,SoejimaParkerBultinckEtAl2020,XieMacDonald2020,WuSarma2020,DasLuHerzog-Arbeitman2021,BernevigSongRegnaultEtAl2021,LiuKhalafLee2021,SaitoGeRademaker2021,JiangLaiWatanabe2019,PotaszXieMacDonald2021,FaulstichStubbsZhuEtAl2023,BeckerLinStubbs2023}.

These models have generally not systematically modeled the effects of structural relaxation. Structural relaxation is the process by which atoms re-arrange themselves in order to minimize their total mechanical energy, taking into account both the monolayer stiffness and energy differences between interlayer stacking configurations \cite{Cazeaux_Clark_Engelke_Kim_Luskin_2023,Carr_Massatt_Torrisi_Cazeaux_Luskin_Kaxiras_2018,Yoo_Engelke_Carr_Fang_Zhang_Cazeaux_Sung_Hovden_Tsen_Taniguchi_etal_2019}. This effect is known to occur in TBG and to significantly modify its single-particle electronic properties (band structure) \cite{Fang_Carr_Zhu_Massatt_Kaxiras_2019,relaxedblg23,Yoo_Engelke_Carr_Fang_Zhang_Cazeaux_Sung_Hovden_Tsen_Taniguchi_etal_2019, Kang_Vafek_2023, Vafek_Kang_2023}, and is expected to affect the many-body phase diagram as well \cite{relaxedheavyfermion}.

The effect of relaxation on the single-particle electronic properties of TBG has often been modeled by varying a single parameter in the Bistritzer-MacDonald model: the ratio of the interlayer hopping strengths between graphene sublattices of the same type (A or B) and of different types. This is natural, since relaxation tends to enlarge areas of (energetically favorable) AB and BA stacking and shrink areas of (unfavorable) AA and BB stacking, so that the hopping energy between distinct sublattices is relatively enhanced \cite{Carr_Massatt_Torrisi_Cazeaux_Luskin_Kaxiras_2018,Cazeaux_Clark_Engelke_Kim_Luskin_2023,relaxhott23}. It is known, however, that tuning this parameter alone cannot be expected to capture all of the effects of relaxation. For example, it is known that interlayer hopping in relaxed TBG is also much longer range than in the original Bistritzer-MacDonald model \cite{relaxedblg23}. 

The present work introduces a novel reduced-order model of TBG's many-body electronic properties which systematically accounts for the effects of structural relaxation. We do this by first modeling relaxation by coupling linear elasticity to a stacking penalty energy computed from Density Functional Theory (DFT) following \cite{Cazeaux_Luskin_Massatt_2020,Carr_Massatt_Torrisi_Cazeaux_Luskin_Kaxiras_2018,zhu2019moir,Cazeaux_Clark_Engelke_Kim_Luskin_2023,dai2016structure,dai2016twisted}. We then incorporate the effects of relaxation into a single-particle electronic tight-binding model. We then simplify this tight-binding model by focusing on wave-packets spectrally concentrated at the monolayer Dirac points in order to derive a moir\'e-scale continuum PDE model which we refer to as the relaxed BM model. We finally add a Coulomb electron-electron interaction and project both parts of the many-body Hamiltonian down to the relaxed BM model's flat bands.
We recall that graphene has two inequivalent Dirac points, denoted \(\bvec{K}\) and \(\bvec{K}'\), which are related to each other by time reversal symmetry.
  While these two Dirac points, referred to as ``valleys'', are far apart in momentum space, previous numerical studies \cite{BultinckKhalafLiuEtAl2020} have shown that including wavepackets centered at both \(\bvec{K}\) and \(\bvec{K}'\) has a profound impact on the nature of the many-body ground states.
  In fact, recent experiments have shown states which hybridize the two valleys are energetically favored \cite{NuckollsLeeOhEtAl2023}.

The other main contribution of the present work is to compute Hartree-Fock minimizers for the new model and compare these results with computations of Hartree-Fock minimizers for the model where the relaxed BM model is replaced by the ordinary BM model with relaxation accounted for by varying the ratio of AA to AB hopping. Following \cite{FaulstichStubbsZhuEtAl2023}, we initialize our Hartree-Fock calculations in the chiral limit (BM model with ratio of AA to AB hopping set $= 0$), where the ground state manifold is analytically known. We then interpolate between the chiral limit and the limiting model of interest: either our relaxed BM model, or for comparison, the BM model with a realistic value for the ratio of AA to AB hoppings. As we do this, we track important observables such as the HOMO-LUMO gap and symmetries of the ground state wavefunctions, finding significant differences between the results for each model. 

The results of the present work clarify the effects of relaxation beyond varying the ratio of AA to AB interlayer hopping on TBG's many-body phase diagram. In particular, our results predict symmetry-breaking phase transitions when the atomic structure is very close to the structure predicted by our relaxation model. In future work, we will investigate the effects of additional strain and model the spin degree of freedom. 

\subsection{Structure of paper}

In \cref{sec:theory}, we introduce our model of mechanical relaxation in TBG, our electronic tight-binding model of TBG, how these models couple, and then our interacting model. In \cref{sec:an-interacting-model} we describe our Hartree-Fock computations and how we track symmetries of minimizers in detail. In \cref{sec:numerics} we present and describe our numerical results. We conclude in \cref{sec:conclusion} where we again summarize our results and describe some future perspectives. 




\section{Theory: Single-particle Model} \label{sec:theory}

We will first establish the necessary notation for describing the geometry of TBG \cite{2009Castro-NetoGuineaPeresNovoselovGeim,Watson_Luskin_2021,Watson_Kong_MacDonald_Luskin_2023,Cazeaux_Clark_Engelke_Kim_Luskin_2023}. The triangular lattice of a single sheet of graphene can be described by a fundamental matrix $A$, whose columns are the primitive lattice vectors
\begin{equation}
    A := (\bvec a_1, \bvec a_2), \quad \bvec a_1 := \frac{a}{2}(1,\sqrt{3})^\top, \quad \bvec a_2 := \frac{a}{2}(-1,\sqrt{3})^\top, 
\end{equation}
where $a\approx 2.5 \text{ \AA}$ is the lattice constant. The graphene Bravais lattice $\mathcal{R}$ and a unit cell $\Gamma$ can be defined as
\begin{equation}
    \mathcal{R} := \{ \bvec R = A\bvec n: \bvec n\in \mathbb{Z}^2 \}, \quad 
    \Gamma=\{ A \alpha : \alpha \in [0,1)^{2} \}.
\end{equation}
Within a unit cell indexed by $\bvec R$, there are two atoms at physical locations $\bvec R + \vec \tau^\A$ and $\bvec R + \vec\tau^\B$, which we define as
\begin{equation} \label{eq:relative_shift}
\vec\tau^\A := (0, 0)^\top, \quad \vec\tau^\B : = \left(0, \delta \right)^\top, \quad \delta:= \frac{a}{\sqrt{3}}.
\end{equation}
These atoms are in sub-lattices A and B respectively, denoted by $\sigma,$ and the relative shift between two sub-lattices $\delta$ is also the minimum distance between two atoms in the same layer. 

TBG is obtained by stacking two identical layers with a relative interlayer twist. Suppose the relative twist angle is $\theta$, then the lattices are denoted by 
\begin{equation}
	 A_j = \rot j A, \quad \vec \tau_j^\sigma  = \rot j \vec \tau^\sigma, \quad \theta _1 = -\frac{\theta}{2}, \quad \theta_2 = \frac{\theta}{2}, \quad \sigma\in \{\A,\, \B\},
\end{equation}
where $\mathfrak R(\theta) $ denoted the counterclockwise rotation by $\theta.$
The rotated Bravais lattice and unit cell are similarly defined
\begin{equation}
	\mathcal R_j := \{\bvec R_j = A_j \bvec n : \bvec n \in \mathbb Z^2\}, \quad \Gamma_j := \{A_j \alpha : \alpha \in [0,1)^2\}.
	\end{equation}	
The moir\'e pattern emerges when two slightly mismatched layers are stacked. It can be defined precisely as the lattice of periodicity for a continuously interpolated interlayer disregistry; see \cite{Nam_Koshino_2017}, for example. The moir\'e lattice vectors, lattice, and fundamental cell are defined by
\begin{equation}
	A_\m := \left(A_1^{-1} - A_2^{-1}\right)^{-1}, \quad \mathcal R_\m := \left\{ \bvec{R}_\m = A_\m \bvec{n} : \bvec{n} \in \mathbb{Z}^2 \right\}, \quad \Gamma_\m := \{A_\m \alpha :  \alpha \in [0,1)^2\}.
\end{equation}
 The moir\'e reciprocal lattices and fundamental cell (Brillouin zone) are defined through the dual relation $A_\m^T B_\m = 2\pi I$, so
\begin{equation} \label{eq:moire_recip}
	B_\m := 2\pi A_\m ^{-T}, \quad \mathcal R_\m^* := \left\{ \bvec{G}_\m = B_\m \bvec{n} : \bvec{n} \in \mathbb{Z}^2 \right\}, \quad  \Gamma_\m^* := \left\{ B_\m \bvec{\beta} : \bvec{\beta} \in [0,1)^2 \right\}.
\end{equation}

Important quantities for monolayer graphene are its Dirac points
\begin{equation}
    \bvec{K} := \frac{ 4 \pi }{ 3 a } ( 1, 0 )^\top, \quad \bvec{K}' := - \bvec{K},
\end{equation}
near to which the monolayer dispersion relation is conical, described by an effective Dirac operator \cite{2009Castro-NetoGuineaPeresNovoselovGeim}. After the twist, the Dirac points of each layer move as $\bvec K_1 = \rot 1 \bvec K$, $\bvec K_2 = \rot 2 \bvec K$. We define their midpoint and difference as
\begin{equation}
	\tilde{ \bvec K} := \frac{\bvec K_1 + \bvec K_2}{2}, \quad \bvec s_1 := \bvec K_1 - \bvec K_2.
\end{equation}

\begin{figure}[h]
\centering
\begin{subfigure}{.25\columnwidth}
\includegraphics[width=.9\columnwidth]{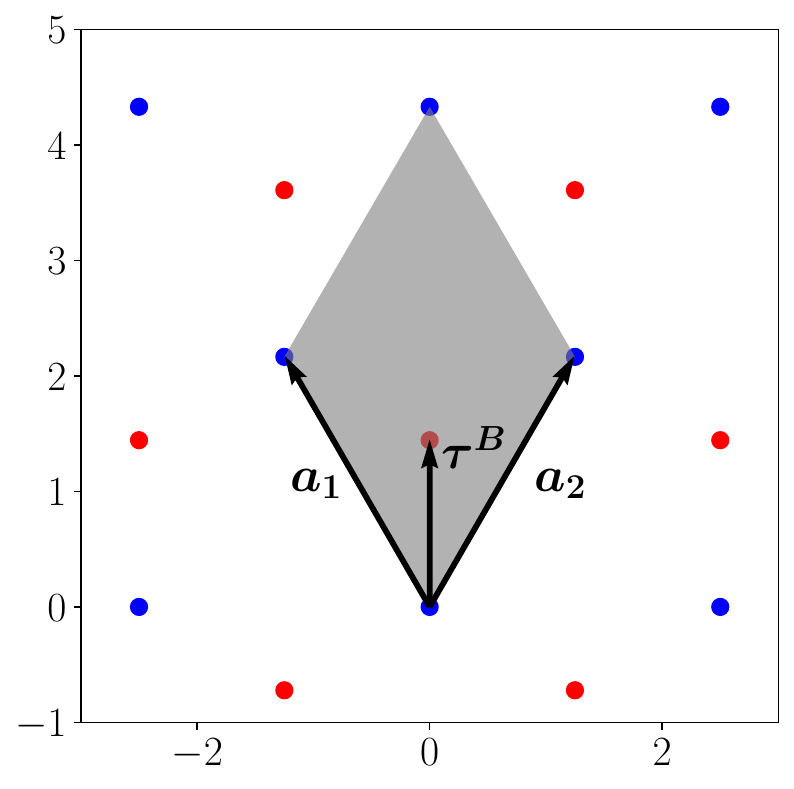}
\caption{\label{fig:graphene}}
\end{subfigure}
\begin{subfigure}{.35\columnwidth}
\includegraphics[width=.9\columnwidth]{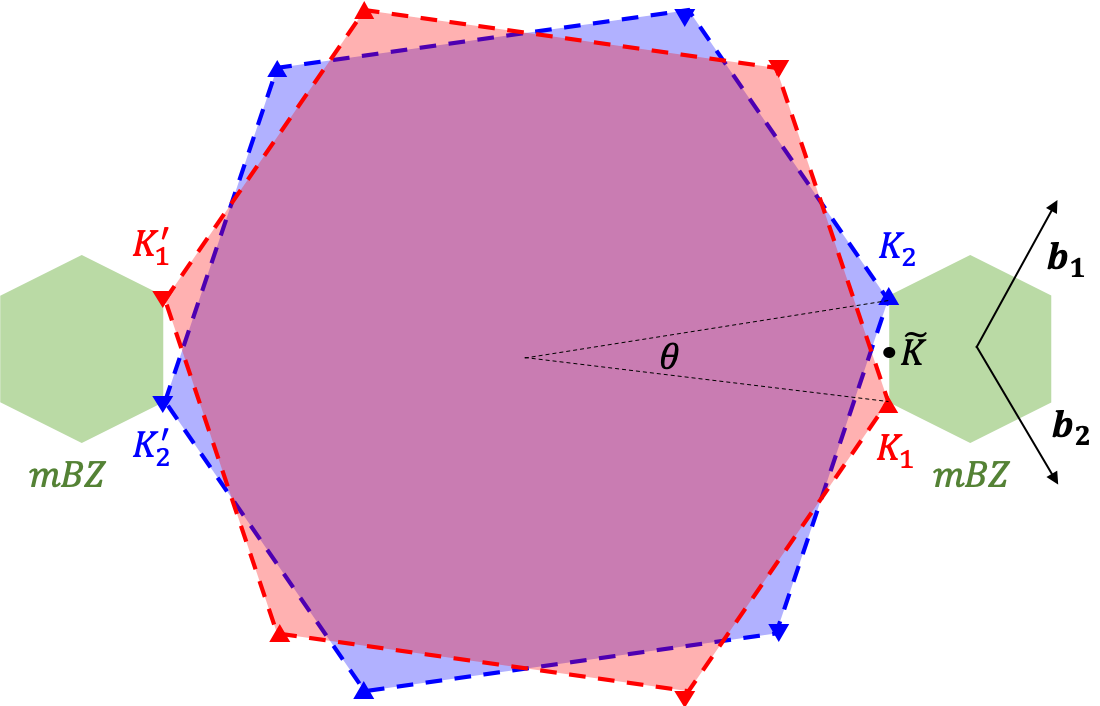}
\caption{\label{fig:mBZ}}
\end{subfigure}
\begin{subfigure}{.35\columnwidth}
\includegraphics[width=.9\columnwidth]{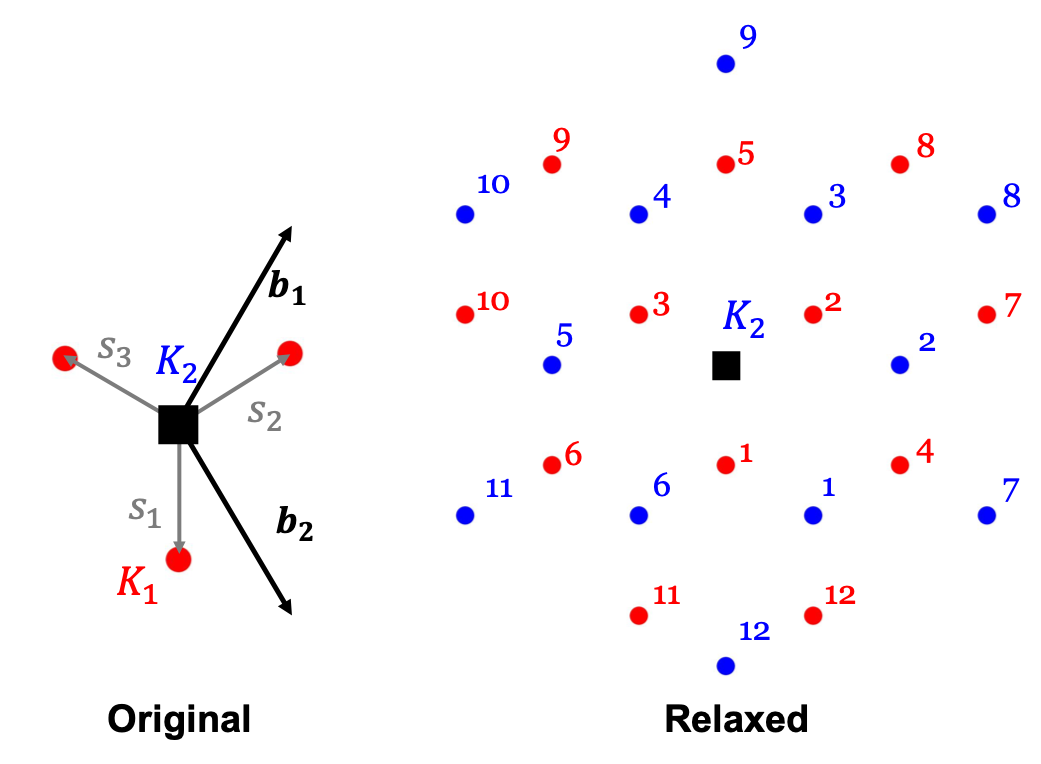}	
\caption{\label{fig:mom_hops}}
\end{subfigure}
\caption{\subref{fig:graphene} Hexagonal graphene lattice, monolayer lattice vectors $\bvec a_1, \bvec a_2$, a unit cell (gray), and the relative shift between A and B sublattices $\vec \tau^B$.  \subref{fig:mBZ} Brillouin zones of individual layers (red and blue), the Dirac points $\bvec K_\ell, \bvec K_\ell'$ and the two moir\'e Brillouin zones around $\bvec K$ and $\bvec K'$ valleys (green) with a set of moir\'e reciprocal lattice vectors $\bvec b_1, \bvec b_2$. We expand our wave function around $\tilde{\bvec K}$. \subref{fig:mom_hops} The momentum hops around $\bvec K_1$ and $\bvec K_2$. The original BM model has only three interlayer momentum hops $\bvec s_1$, $\bvec s_2 = \bvec s_1 + \bvec b_1$ and $\bvec s_3 = \bvec s_1 - \bvec b_2$. The relaxed BM model allows further interlayer hopping with momentum shifts $\bvec s_1 + \mathcal R_\m^*$ (red), and also intralayer hopping with momentum shifts $\mathcal R_\m^*$ (blue). We keep the first three shells of interlayer hopping, and first two shells of intralayer hopping (12 hoppings each).}
\end{figure}

\subsection{Bistritzer-MacDonald Model}\label{sec:bm}

We now introduce the underlying tight-binding model to describe electronic properties of TBG \cite{Watson_Kong_MacDonald_Luskin_2023,momentumspace17,dos17}. We start with the discrete wave function
\begin{equation}
    \Psi = (\Psi_1^\A, \Psi_1^\B, \Psi_2^\A, \Psi_2^\B)^\top, \quad \Psi_\ell^\sigma = (\Psi^\sigma_{\bvec R_\ell})_{\bvec R_\ell \in\mathcal R_\ell} \in \ell^2(\mathcal{R}_\ell).
\end{equation} 
The Hamiltonian acts on the wave functions through
\begin{equation}\label{eq:H_tb}
\begin{split}
(H \Psi)_{\bvec R_1}^\sigma 	 =  \sum_{\bvec R_1'\in\mathcal R_1} \sum_{\sigma'} h_{11}^{\sigma\sigma'}(\bvec R_1 + \vec \tau_1^\sigma - \bvec R_1' -\vec \tau_1^{\sigma'})\Psi_{\bvec R_1'}^{\sigma'} 
  + \sum_{\bvec R_2'\in\mathcal R_2} \sum_{\sigma'} h_{12}^{\sigma\sigma'}(\bvec R_1 + \vec \tau_1^\sigma  - \bvec R_2 - \vec\tau_2^{\sigma'})\Psi_{\bvec R_2}^{\sigma'},
 \end{split}
\end{equation}
where $h_{11}$ and $h_{12}$ are the intralayer and interlayer hopping functions. We use hopping functions based on density functional theory calculations \cite{Fang_Kaxiras_2016}.

\begin{figure}[h!]
\centering
\begin{subfigure}{.3\columnwidth}
\includegraphics[width=\columnwidth]{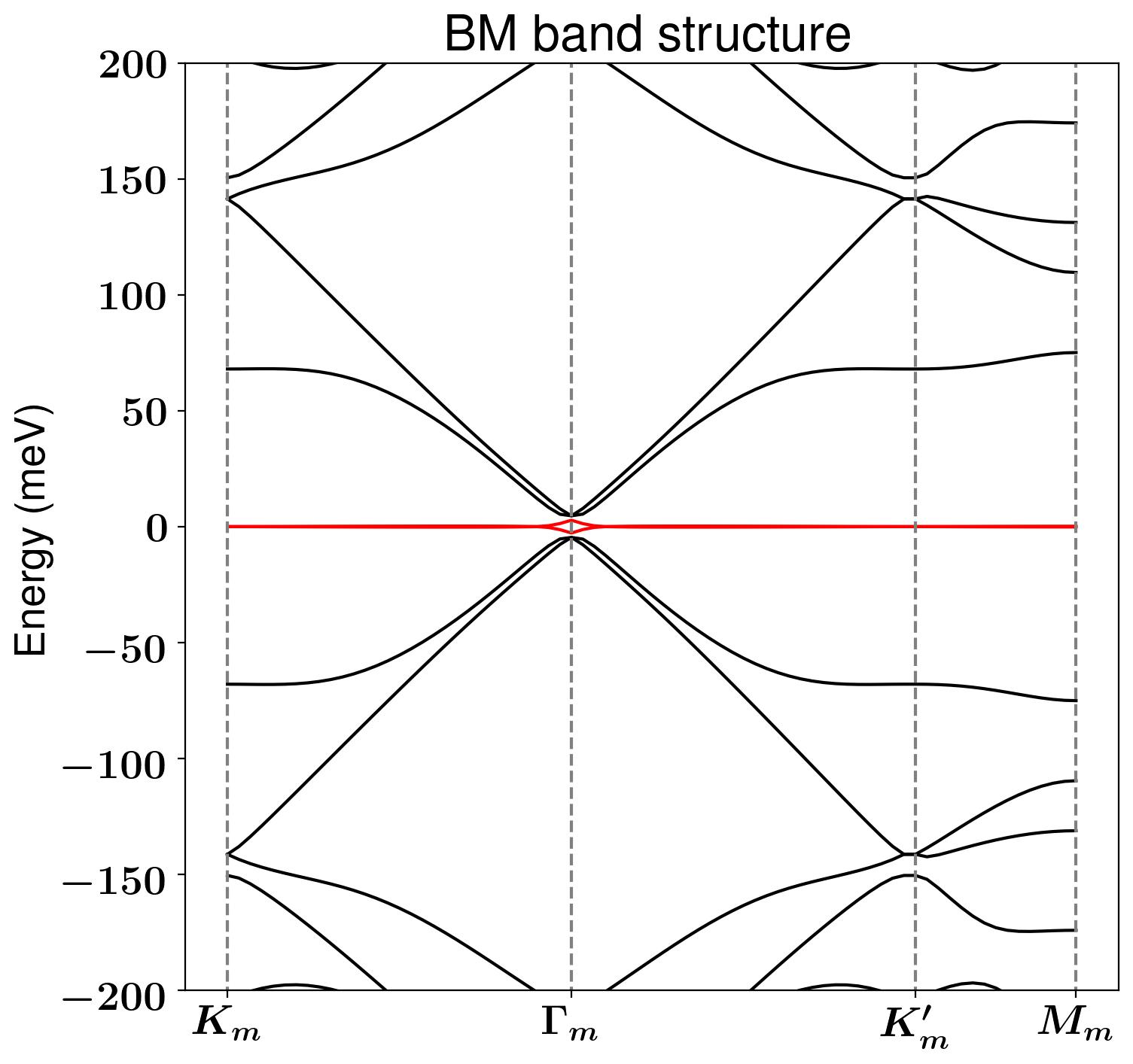}
\caption{\label{fig:BM_bands}}
\end{subfigure}
\begin{subfigure}{.3\columnwidth}
\includegraphics[width=\columnwidth]{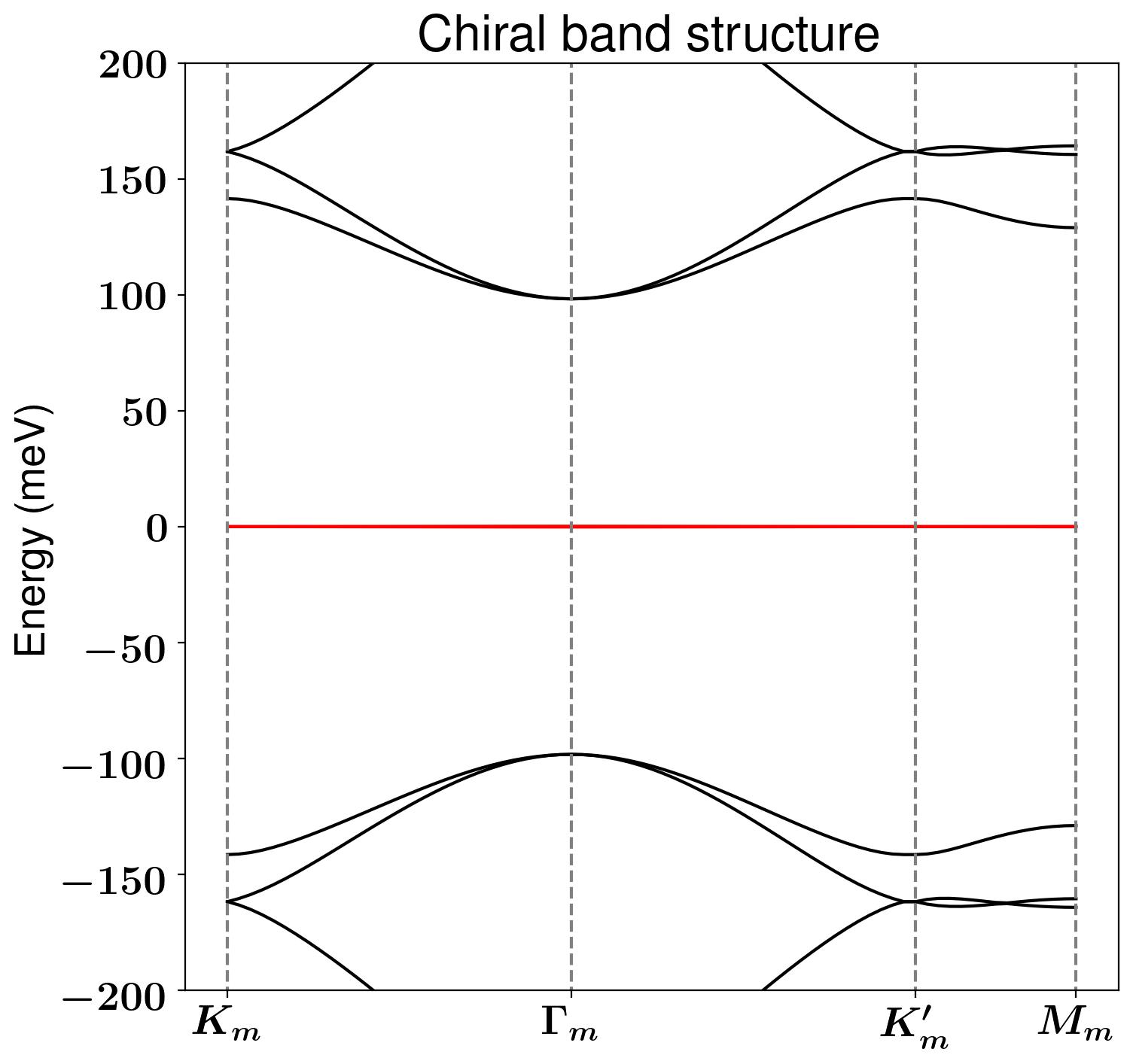}
\caption{\label{fig:chiral_bands}}
\end{subfigure}
\begin{subfigure}{.3\columnwidth}
\includegraphics[width=\columnwidth]{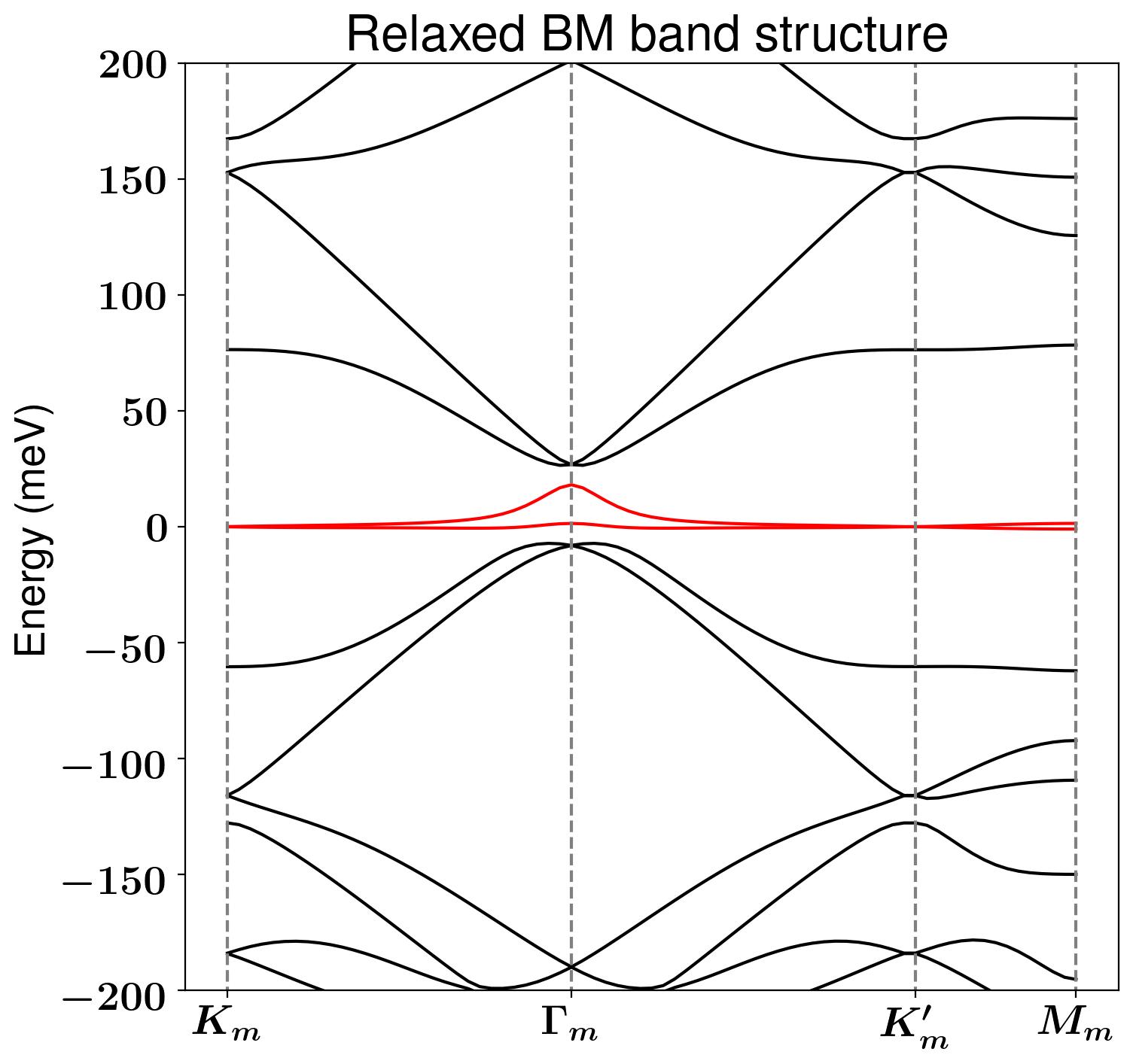}	
\caption{\label{fig:relaxed_BM_bands}}
\end{subfigure}
\caption{Single particle band structures for $\bvec K$ valley of  \subref{fig:BM_bands} BM model, \subref{fig:chiral_bands} Chiral Model, \subref{fig:relaxed_BM_bands} Relaxed BM model at $1.05^\circ$.
    The projected bands are highlighted in red. Note that the gap between the projected bands and the ``remote bands'' is small in the BM model \subref{fig:BM_bands}. To avoid these bands touching we consider $\kappa \leq .95$ in this work. This is justified since relaxation tends to decrease this ratio significantly; in our relaxed Bistritzer-MacDonald model we calculute this parameter as $\approx .7$, see Appendix \ref{sec:relax-bm-inter}. }
\end{figure}

The Bistritzer-MacDonald (BM) model is a moir\'e-periodic continuum approximation to \eqref{eq:H_tb} which greatly simplifies single-particle electronic property calculations
\cite{Bistritzer_MacDonald_2011}. Previous work of three of the authors proved that at the first magic angle, the BM model is the effective model for wave packets spectrally concentrated around the monolayer Dirac points \cite{Watson_Kong_MacDonald_Luskin_2023,Kong_Liu_Luskin_Watson_2024,Quinn_Kong_Luskin_Watson_2025}. 
We generate the discrete wave function by evaluating a continuous $\psi$ on the lattice sites
\begin{equation}
	\Psi_{\bvec R_\ell}^\sigma = \left. \psi(\bvec r; \sigma, \ell) \right|_{\bvec r = \bvec R_\ell + \vec \tau_\ell^\sigma}, 
\end{equation}
where at this point we adopt a convenient notation where the sublattice and layer degrees of freedom appear as discrete variables. We interpret $\psi(\cdot;\sigma,\ell)$ as describing the electron density on layer $\ell \in \{1,2\}$, sublattice $\sigma \in \{\A, \B\}$, and in valley $\bvec K$.
%
%
The BM model, the Hamiltonian which acts on $\psi$, in real space is \begin{equation} \label{eq:H_BM}
   \Hbm = 
    \begin{pmatrix} v \vec{\sigma} \cdot (-i \nabla_{\bvec r} - \bvec K_1) &  T(\bvec r) \\  T^\dagger(\bvec r) & v \vec{\sigma}\cdot (-i \nabla_{\bvec r} - \bvec K_2) \end{pmatrix}.
    \end{equation}
We can interpret the Hamiltonian as an expansion around the rotated Dirac point on both layers. 
The interlayer potential terms are
\begin{equation}
\label{eq:T_matrix}
	T(\bvec r) = \sum_{j=0}^2 T_j e^{-i \bvec Q_j \cdot \bvec r}, 
	\quad T_j = 
	\begin{pmatrix}
		w_0 & w_1 e^{-i 2\pi j/3} \\ w_1 e^{i 2\pi j/3} & w_0
	\end{pmatrix},
\end{equation}
which depends on moir\'e reciprocal lattice vectors $\bvec Q_j.$ Specifically, $\bvec Q_0 = 0$, $\bvec Q_1 = \bvec b_1$ and $\bvec Q_2 = -\bvec b_2$, where $\vec b_j$ are the columns of the matrix $B_\m$ appearing in \eqref{eq:moire_recip}, and such that the momentum shifts between layers are $\bvec s_1,\bvec s_2$ and $\bvec s_3$, see the left panel of \cref{fig:mom_hops}. The ratio $\kappa = w_0/ w_1$ is commonly used to model relaxation effects, where AB stacking is preferred over AA stacking configurations. In the present work, we restrict attention to ratios $0 \leq \kappa \leq 0.95$ to ensure the flat bands are separated, see \cref{fig:BM_bands}. This is reasonable since estimates of the true value of $\kappa$ generally put it well within this interval; when we construct our relaxed Bistritzer-MacDonald model we find $\kappa \approx .7$ (see Section \ref{sec:model_relax} and Appendix \ref{sec:relax-bm-inter}). 

Using Bloch's theorem, we can label the eigenfunctions of the BM Hamiltonian using momentum $\bvec k \in \Gamma^*_\m$ and band index $\eta \in \mathbb Z$
\begin{equation}
  \label{eq:bm_bloch}
  \Hbm \psi_{\eta \bvec k}(\bvec r) = \eps_{\eta \bvec k} \psi_{\eta \bvec k}(\bvec r). 
\end{equation}
To symmetrize the expression, we put the momentum space origin at the midpoint of rotated Dirac points between layers $\tilde {\bvec{K}} = (\bvec K_1 + \bvec K_2) / 2$, and seek eigenfunctions with the form
	\begin{equation}\label{eq:ansatz}
    \psi_{\eta \bvec k}(\bvec r; \sigma, \ell) = e^{i \tilde {\bvec K} \cdot \bvec r} e^{i\bvec k \cdot \bvec r} u_{\eta \bvec k}(\bvec r; \sigma, \ell) = \frac{1}{\left|\Gamma_\m^*\right|} \sum_{\bvec G \in \mathcal R_\m^*}e^{i(\tilde {\bvec K}  + \bvec k + \bvec G)\cdot \bvec r} {u _{\eta \bvec k}}(\bvec G; \sigma, \ell).
\end{equation}
The eigenfunctions contain a fast oscillating plane wave $e^{i \tilde {\bvec K} \cdot \bvec r}$ and a moir\'e scale smooth envelope function.
Here $u_{\eta \bvec k}(\bvec r)$ is periodic with respect to the moir\'e lattice $\mathcal R_\m$, that is $u_{\eta \bvec k}(\bvec r; \sigma, \ell) = u_{\eta \bvec k}(\bvec r + \bvec a; \sigma, \ell)$ for $\bvec a \in \mathcal R_\m$. With some abuse of notation, $u_{\eta \bvec k}(\bvec r;\sigma,\ell)$ and $u_{\eta \bvec k}(\bvec G;\sigma,\ell)$ represents the coefficients of the eigenfunctions in real space and in momentum space.
Subsequently, we can define Bloch-Floquet transformed Hamiltonian
\begin{equation}
	\left[\Hbm(\bvec k) u_{\eta \bvec k}\right](\bvec G) = \sum_{\bvec G' \in \mathcal R^*_\m} \left[ \Hbm(\bvec k) \right] _{\bvec G, \bvec G'} u_{\eta \bvec k}(\bvec G'),
\end{equation}
\begin{equation}\label{eq:H_BM_k}
	  \left[\Hbm (\bvec k)\right]_{\bvec G, \bvec G'} = 
	 \begin{pmatrix} 
	 \displaystyle v \vec{\sigma} \cdot \left(\bvec k + \frac{\bvec s_1}{2}+ \bvec G\right) \delta_{\bvec G, \bvec G'} & 
	 \displaystyle \sum_{j=0}^2 T_j \delta_{\bvec G, \bvec G' +\bvec Q_j} \\  
	 \displaystyle \sum_{j=0}^2 T_j^\dagger \delta_{\bvec G, \bvec G' -\bvec Q_j} & 
	 \displaystyle v \vec{\sigma}\cdot \left(\bvec k - \frac{\bvec s_1}{2}+ \bvec G\right) \delta_{\bvec G, \bvec G'}  
	 \end{pmatrix},
\end{equation}
where the momentum hops $\bvec Q_j$ are the first shell of moir\'e reciprocal lattices  (see \cref{fig:mom_hops}). 

The chiral model is generated by setting $w_0 = 0$ in \cref{eq:T_matrix} \cite{Tarnopolsky_Kruchkov_Vishwanath_2019}.  The chiral model has exact flat bands (see \cref{fig:chiral_bands}), its eigenvalues satisfy $\eps_{\eta \bvec k} = 0$ for $k\in \Gamma_\m^*$ and $\eta = \pm 1$ \cite{Watson_Luskin_2021,Tarnopolsky_Kruchkov_Vishwanath_2019,Zworskimagic22,Zworskispectra21l} The significance of the chiral model is that after adding electron-electron interactions, the ground states of the resulting interacting chiral model can be solved exactly \cite{BeckerLinStubbs2023,Stubbs_Becker_Lin_2024,StubbsRagoneMacDonaldEtAl2025}.

The previous expansion is around the $\bvec K$ valley only. To introduce a valleyful single-particle Hamiltonian, we can rewrite the wave packet ansatz as 
\begin{equation}\label{eq:ansatz_2V}
	\Psi_{\bvec R_\ell}^\sigma =  \left. \psi(\bvec r; \sigma, \ell, \bvec{K}) \right|_{\bvec r = \bvec R_\ell + \vec \tau_\ell^\sigma} +  \left. \psi(\bvec r; \sigma, \ell, \bvec{K}') \right|_{\bvec r = \bvec R_\ell + \vec \tau_\ell^\sigma},
\end{equation}
where we have introduced the notation $\psi(\vec{r};\sigma,\ell,\nu)$ where $\nu \in \{\bvec{K},\bvec{K}'\}$ is the valley degree of freedom. The plane waves $e^{i \tilde {\bvec K} \cdot \bvec r}$ and $e^{-i \tilde {\bvec K} \cdot \bvec r}$ are orthogonal in $\ell^2$, and the envelope functions are slowly varying, so we can approximate the valleyful dynamics with a decoupled Hamiltonian. The valleyful Hamiltonian is the direct sum of $\bvec K$ and $\bvec K'$ valley components acting on $\psi$
\begin{equation}
\begin{split}
	\HbmVV & := \Hbm \oplus \Hbm'\\ 
	& =\begin{pmatrix}   v \vec{\sigma} \cdot (-i \nabla_{\bvec r} - \bvec K_1) &  T(\bvec r) &  & \\  T^\dagger(\bvec r) & v \vec{\sigma}\cdot (-i \nabla_{\bvec r} - \bvec K_2)  &   &  \\ &   & - v \overline{\vec{\sigma}} \cdot (-i \nabla_{\bvec r} - \bvec K_1') &  \overline {T(\bvec r)} \\ & &  \overline {T^\dagger(\bvec r)} & - v \overline {\vec{\sigma}} \cdot (-i \nabla_{\bvec r} - \bvec K_2') \end{pmatrix}.
\end{split}
\end{equation}
The blocks $\Hbm$ and  $\Hbm'$ are related through a time-reversal symmetry (see \cref{sec:symm-single-part} for a detailed discussion). We now seek eigenfunctions of the form
\begin{equation}\label{eq:eigenfunction_new}
\begin{split}
    \psi_{\eta \bvec k}(\bvec r; \sigma, \ell, \bvec{K}) &= e^{i (\tilde{\bvec K} + \bvec k) \cdot \bvec r}  u_{\eta \bvec k}(\bvec r; \sigma, \ell, \bvec{K}), \\ \psi_{\eta \bvec k}(\bvec r; \sigma, \ell, \bvec{K}') &= e^{- i (\tilde{\bvec K} + \bvec k) \cdot \bvec r}  u_{\eta \bvec k}(\bvec r; \sigma, \ell, \bvec{K}'), \end{split} \quad \eta \in \mathbb Z, \, \bvec k \in \Gamma_\m^*.
\end{equation}
Note that $\bvec{k}$ in \eqref{eq:eigenfunction_new} is a relative wave-vector, in contrast to the total wave-vectors $\pm (\tilde{\bvec{K}}+\bvec{k})$. Note also that we adopt the convention that relative wave-vectors in the $\bvec{K}'$ valley have the opposite orientation to those in the $\bvec{K}$ valley. This convention will turn out to be natural when we discuss the 1-particle reduced density matrix in Section \ref{sec:hartree-fock-theory}. Since $\HbmVV$ is block diagonal, we can define the eigenfunctions that are non-zero only on one of the valley indices
\begin{equation}
	\psi_{(\nu, \eta) \bvec k}(\bvec r; \sigma, \ell) := \psi_{\eta \bvec k}(\bvec r; \sigma, \ell, \nu), \quad u_{(\nu, \eta) \bvec k}(\bvec r; \sigma, \ell) := u_{\eta \bvec k}(\bvec r; \sigma, \ell, \nu),
\end{equation}
and the tuples $(\nu, \eta)$ denote the generalized index that incorporates valley and band degrees of freedom. In this notation, Bloch-Floquet transformed Hamiltonian $\Hbm(\bvec k)$ acts on $u_{(\bvec K, \eta)\bvec k}$, and $\Hbm'(\bvec k)$ acts on $u_{(\bvec K', \eta)\bvec k}$. The time-reversal symmetry for the Hamiltonians is then $\Hbm'(\bvec k)_{\bvec G,\bvec G'} = \overline{\Hbm (\bvec k)_{-\bvec G,-\bvec G'}}$.


\subsection{Modeling Structural Relaxation}
\label{sec:model_relax}

\subsubsection{Elastic Model for Relaxation}
\begin{figure}[h]
\centering
\begin{subfigure}{.4\columnwidth}
\includegraphics[width=.8\columnwidth]{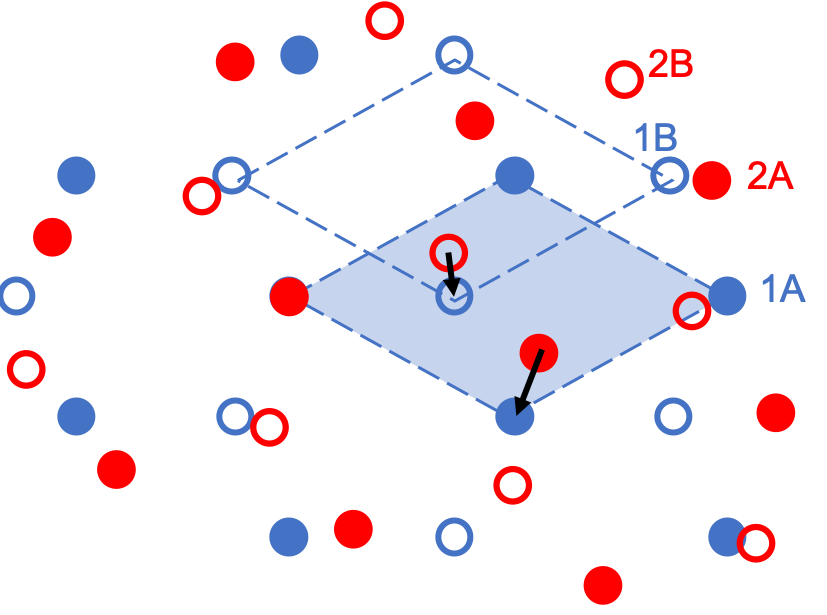}
\caption{\label{fig:disregistry}}
\end{subfigure}
\begin{subfigure}{.4\columnwidth}
\includegraphics[width=\textwidth]{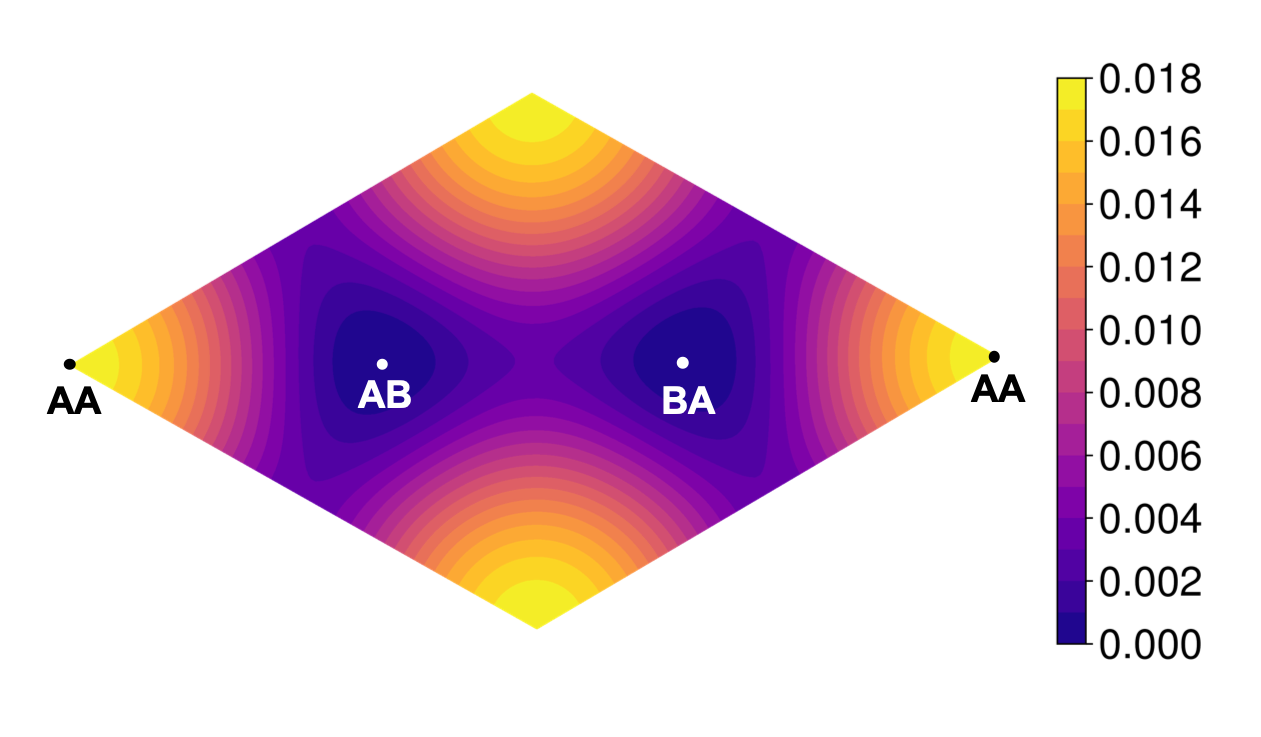}
\caption{\label{fig:GSFE}}
\end{subfigure}
\caption{\subref{fig:disregistry} Illustration of local stacking configuration of layer 2 (red) with respect to layer 1 (blue). The disregistry is a vector in the unit cell $\Gamma_1$ of layer 1. We also account for the sublattice shift when computing the disregistry, atoms in A sublattice are compared to A sublattice of the other layer.  \subref{fig:GSFE} GSFE as a function of disregistry. It is minimized at AB configuration, and maximized at AA configuration. }
\end{figure}

We summarize the approach to modeling mechanical relaxation in TBG. We assume the relaxation can be described by a smooth, moir\'e-periodic displacement field $\bvec U_{\m, \ell}$, where $\ell$ denotes the number of layer. 
Following the formulation in \cite{Carr_Massatt_Torrisi_Cazeaux_Luskin_Kaxiras_2018,Cazeaux_Luskin_Massatt_2020,Cazeaux_Clark_Engelke_Kim_Luskin_2023,zhu2019moir}, we assume the displacement functions minimize the energy functional
 \begin{equation}\label{eq:relax_optim}
	\mathcal E [\bvec U_{\m,1}, \bvec U_{\m, 2}] := \sum_{\ell=1}^2 \mathcal E^\text{intra}_\ell(\bvec U_{\m,\ell}) + \mathcal E^\text{inter}(\bvec U_{\m,1} -  \bvec U_{\m, 2}).
\end{equation}
The intralayer energy is the linear elastic energy of deforming the layers
\begin{equation}
	\mathcal E^\text{intra}_\ell(\bvec U) := \int_{\Gamma_\m} \frac{\lambda}{2} (\operatorname{div} \bvec U )^2 + \mu\eps(\bvec U) : \eps(\bvec U) \dee \bvec x, \quad \eps(\bvec U) := \frac{1}{2}\left(\nabla \bvec U + \nabla \bvec U ^\top \right),
\end{equation}
where $\lambda, \mu>0$ are the Lam\'e parameters of single layer graphene. 
The interlayer energy measures the energy cost of not aligning in a Bernal-stacked (AB or BA everywhere) configuration. We use the generalized stacking fault energy (GSFE), which provides the energy landscape of a unit cell that depends only on the relative stacking between the two layers. The interlayer energy is
\begin{equation}
	\mathcal E^\text{inter}(\bvec V) := \frac{1}{2}\int_{\Gamma_\m} \Phi_1(\vec \gamma_2(\bvec x) + \bvec V(\bvec x)) + \Phi_2(\vec\gamma_1(\bvec x) - \bvec V(\bvec x)) \dee \bvec x,
\end{equation}
where the GSFE energy $\Phi_1$ is periodic on $\Gamma_2$, and $\vec\gamma_2 : \Gamma_\m \to \Gamma_2$, $\vec\gamma_2(\bvec x) = (I - A_2A_1^{-1})\bvec x$ is the initial local stacking configuration of layer 1 with respect to layer 2 (see \cref{fig:disregistry}). The interlayer energy depends only on the relative deformation of the lattices. The energy landscape of $\Phi_j$ shows the stacking energy is maximized when the local configuration is at AA, and minimized at AB (see \cref{fig:GSFE}).

The final position of the relaxed atoms is
\begin{equation}\label{eq:displacement}
	\bvec R_1 + \vec \tau_1^\sigma \mapsto \bvec R_1 + \vec \tau_1^\sigma + \bvec U_{\m,1}(\bvec R_1 + \vec \tau_1^\sigma), \quad 
	\bvec R_2 + \vec \tau_2^\sigma \mapsto \bvec R_2 + \vec \tau_2^\sigma + \bvec U_{\m,2}(\bvec R_2 + \vec\tau_2^\sigma),
\end{equation}
and since the graphene layers are initially identical, we have $\bvec U_{\m,2}=-\bvec U_{\m,1}$.

\begin{figure}[h]
\centering
\includegraphics[width=.6\textwidth]{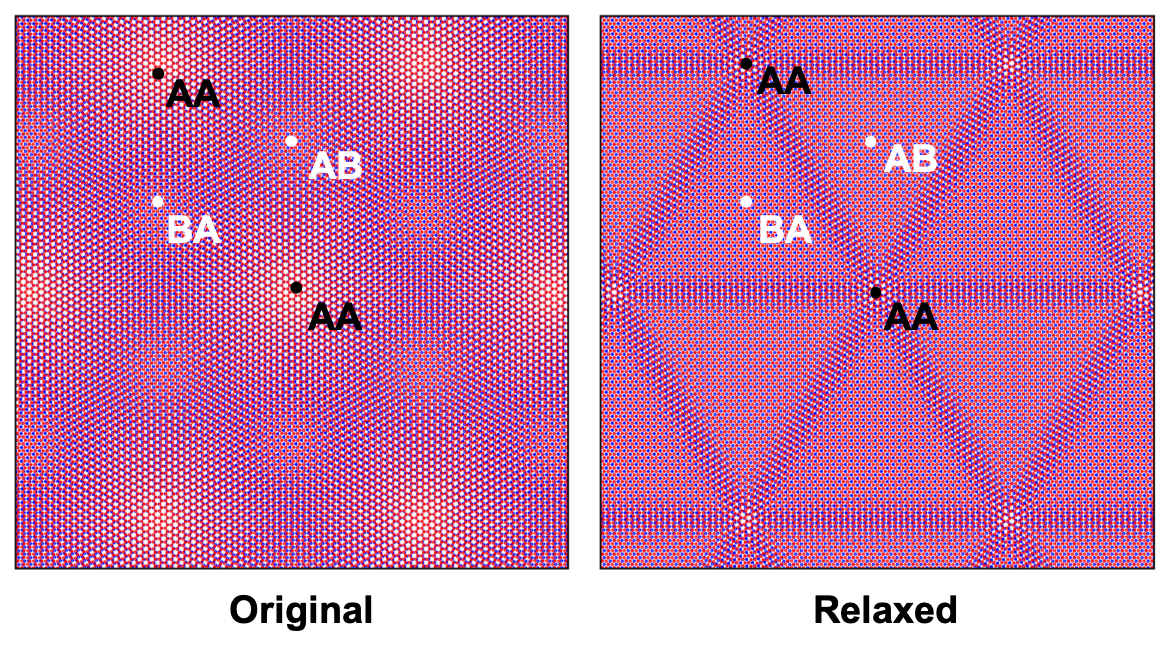}	
\caption{Original and relaxed atom configuration for TBG at $\theta = 1.5^\circ$. To generate the relaxed figure, the GSFE was scaled to be 50 times stronger to emphasize the triangular pattern after relaxation. Regions of AA stacking shrink in size, while AB and BA stacking regions grow.}\label{fig:relax_pos}
\end{figure}

\subsubsection{Relaxed BM Model}
\label{sec:relax-bistr-macd}
The relaxed tight-binding Hamiltonian takes into account the relaxed position of the lattice sites 
\begin{equation}\label{eq:H_tb_relax}
\begin{split}
(H \Psi)_{\bvec R_1}^\sigma 	 = & \sum_{\bvec R_1'\in\mathcal R_1} \sum_{\sigma'} h_{11}^{\sigma\sigma'}(\bvec R_1 + \vec \tau_1^\sigma + \bvec U_{\m,1}(\bvec R_1 + \vec \tau_1^\sigma ) - \bvec R_1' -\vec \tau_1^{\sigma'} - \bvec U_{\m,1}(\bvec R_1' + \vec \tau_1^{\sigma'}))\Psi_{\bvec R_1'}^{\sigma'} \\
 & + \sum_{\bvec R_2'\in\mathcal R_2} \sum_{\sigma'} h_{12}^{\sigma\sigma'} (\bvec R_1 + \vec \tau_1^\sigma + \bvec U_{\m,1}(\bvec R_1+ \vec \tau_1^\sigma ) - \bvec R_2 - \vec\tau_2^{\sigma'} - \bvec U_{\m,2}( \bvec R_2 + \vec\tau_2^{\sigma'}))\Psi_{\bvec R_2}^{\sigma'}.
 \end{split}
\end{equation}

Using a similar systematic multiscale analysis, we derive a single valley continuum BM model for the relaxed Hamiltonian \cite{Kong2025}.
\begin{equation}
	 \left[\Hrel (\bvec k)\right]_{\bvec G, \bvec G'} = 
	 \begin{pmatrix} 
\begin{split}
	 &D^{(1)}\left(\bvec k + \frac{\bvec s_1}{2}+ \bvec G\right)\delta_{\bvec G, \bvec G'} \\  & \quad + \displaystyle \sum_{j=1}^{12} A^{(1)}_{j}\left(\bvec k + \frac{\bvec s_1}{2}\right)\delta_{\bvec G,\bvec G' + \bvec P_j}
\end{split} &	 
\displaystyle \sum_{j=1}^{12} \tilde T_{j}(\bvec k)\delta_{\bvec G ,\bvec G'+ \bvec Q_j} \\  
	 \displaystyle \sum_{j=1}^{12} \tilde T_{j}^\dagger(\bvec k)\delta_{\bvec G , \bvec G'- \bvec Q_j} & 
\begin{split}
&D^{(2)} \left(\bvec k - \frac{\bvec s_1}{2}+ \bvec G\right) \delta_{\bvec G, \bvec G'} \\ &\quad + \displaystyle \sum_{j=1}^{12} A^{(2)}_{j} \left(\bvec k - \frac{\bvec s_1}{2}\right) \delta_{\bvec G,  \bvec G' + \bvec P_j} 
\end{split}
\end{pmatrix}.
\end{equation}
The detailed form of this operator is given in \cref{sec:relax-bm-intra,sec:relax-bm-inter}. Here, we just remark that $D^{(\ell)}(\bvec k)$ is the Dirac operator with second order corrections, $\bvec P_j$ are the first and second shells of moir\'e reciprocal lattices for intralayer momentum hops (blue in \cref{fig:mom_hops}), and $\bvec Q_j$ are the first three shells of moir\'e reciprocal lattices for interlayer momentum hops (red in \cref{fig:mom_hops}). The intralayer and interlayer scattering terms are of the form
\begin{equation}\label{eq:relaxed_hopping}
	A_j^{(\ell)}(\bvec k) = A_j^{(\ell)} + A_{j,\nabla}^{(\ell)} \cdot \bvec k, \quad  \tilde T_j(\bvec k) = \tilde T_j + \tilde T_{j,\nabla}\cdot \bvec k.
\end{equation}
The relaxed BM model introduces intralayer relaxation effects, longer range momentum scattering, and $\bvec k$-dependent non-local approximations to accurately consider the effect of mechanical relaxation.
The hopping matrices \cref{eq:relaxed_hopping} 
are computed from the actual position of the relaxed lattices, and account for the preference of AB to AA stacking. For more details, see \cref{sec:relax-bm-intra,sec:relax-bm-inter}.

We define the valleyful relaxed BM Hamiltonian similar to original valleyful BM Hamiltonian introduced in \cref{sec:bm}.
%

\subsection{Symmetries of the Single Particle Hamiltonian}
\label{sec:symm-single-part}
When including the valley degree of freedom, the single particle Hamiltonian for twisted bilayer graphene can be written as an \(8 \times 8\) matrix.
These \(8\) entries are indexed by a vector of three quantities \((\sigma, \ell, \nu)\) where \(\sigma\) denotes sublattice, \(\ell\) denotes layer, and \(\nu\) denotes valley.
Since each of these quantities take on two possible values, we can define a family of Pauli operators for each of these degrees of freedom.
These Pauli operators will be useful for expressing the symmetries of the single particle Hamiltonian.

It is notationally simplest to temporarily map each of the discrete degrees of freedom $\sigma, \ell, \nu$ into $\{ \pm 1 \}$. Specifically, we will replace 
\begin{equation}
    \sigma \in \{ \A, \B \} \rightarrow \{1,-1\}, \quad \ell \in \{1,2\} \rightarrow \{1,-1\}, \quad \nu \in \{\bvec{K},\bvec{K}'\} \rightarrow \{1,-1\}.
\end{equation}
We then define the matrices \(\sigma_{x}, \sigma_{y}, \sigma_{z}\) which transform the sublattice indices as follows:
\begin{equation}
  \renewcommand{\arraystretch}{1.3}
  \begin{array}{l}
    \sigma_{x} f(\bvec{r}; \sigma, \ell, \nu) = f(\bvec{r}; -\sigma, \ell, \nu),  \\
    \sigma_{y} f(\bvec{r}; \sigma, \ell, \nu) = (i \sigma) f(\bvec{r}; -\sigma, \ell, \nu) ,\\
    \sigma_{z} f(\bvec{r}; \sigma, \ell, \nu) = \sigma f(\bvec{r}; \sigma, \ell, \nu).
  \end{array}
\end{equation}
These operations correspond to the usual Pauli matrices applied to the sublattice degree of freedom namely:
\begin{equation}
  \sigma_{x} =
  \begin{bmatrix}
    & 1 \\
    1 & 
  \end{bmatrix},
  \qquad
  \sigma_{y} =
  \begin{bmatrix}
    & -i \\
    i & 
  \end{bmatrix},
  \qquad
  \sigma_{z} =
  \begin{bmatrix}
    1 & \\
    & -1 
  \end{bmatrix}.
\end{equation}
The matrices \(\ell_{x}, \ell_{y}, \ell_{z}\) and \(\nu_{x}, \nu_{y}, \nu_{z}\) are defined analogously for layer and valley respectively.

We can now define two important symmetry operators: 
\(C_{2z}\), which denotes in-plane rotation by \(180^{\circ}\), and \(\mathcal{T}\), which denotes spinless time reversal symmetry.
The symmetry operators \(C_{2z}\) and \(\mathcal{T}\) act as follows: 
\begin{equation}
  \begin{split}
    C_{2z} f(\bvec{r}; \sigma, \ell, \nu) & := \sigma_{x} f(-\bvec{r}; \sigma, \ell, \nu), \\
    \mathcal{T} f(\bvec{r}; \sigma, \ell, \nu) & :=  \overline{f(\bvec{r}; \sigma, \ell, \nu)}.
  \end{split}
\end{equation}
One can check that in all cases, the valleyful single particle Hamiltonian satisfies
\begin{equation}
    [ C_{2z} \mathcal{T}, \HbmVV ] = 0, \qquad [ \nu_{x} \mathcal{T}, \HbmVV ] = 0, \qquad [\nu_{y} \mathcal{T}, \HbmVV ] = 0.
\end{equation}
Despite the fact that the single particle Hamiltonian commutes with the symmetries \(C_{2z}\mathcal{T}, \nu_{x} \mathcal{T}, \nu_{y} \mathcal{T}\), it is possible for the Hartree-Fock ground states to spontaneously break these symmetries (see \cref{sec:gauge-fixing-init} for more discussion).

\section{Theory: Many-body Model}
\label{sec:an-interacting-model}

To define an interacting model for TBG, we begin by fixing $n_{k_{x}}, n_{k_{y}} \in \mathbb{Z}_{+}$ and define the $n_{k_{x}} \times n_{k_{y}}$ Monkhorst-Pack grid $\mathcal{K}$ as follows: 
\begin{equation}
  \mathcal{K} = \left\{ \frac{i}{n_{k_{x}}} \bvec{b}_{1} + \frac{j}{n_{k_{y}}} \bvec{b}_{2} : 0 \leq i < n_{k_{x}}, 0 \leq j < n_{k_{y}} \right\} \subseteq \Gamma_{\m}^{*}.
\end{equation}
We project the interaction onto the flat bands, so we restrict the band index $\mathcal N := \{(\nu,\eta): \nu\in    \{\bvec K, \bvec K'\}, \eta\in \{-1, 1\} \}$. This restriction is well-motivated as long as there is an energy gap between the flat bands and the remote bands. For each $\bvec{k} \in \mathcal K$ and each $m \in \mathcal{N}$, let $\hat{f}_{m \bvec{k}}^{\dagger}$ and $\hat{f}_{m \bvec{k}}$ denote the creation and annihilation operator which creates or annihilates an electron in the state $\psi_{m \bvec{k}}(\bvec{r})$.
These operators satisfy the canonical anticommutation relations (CAR)
\begin{equation}
  \begin{split}
    \{ \hat{f}_{m \bvec{k}}^{\dagger}, \hat{f}_{n \bvec{k}'}^{\dagger} \} & = \{ \hat{f}_{m \bvec{k}}, \hat{f}_{n \bvec{k}'} \} = 0, \\[1ex]
    \{ \hat{f}_{m \bvec{k}}^{\dagger}, \hat{f}_{n \bvec{k}'} \} & = \delta_{m,n} \delta_{\bvec{k} - \bvec{k}' \in \Gamma_{m}^{*}}.
  \end{split}
\end{equation}
as well as the boundary conditions 
\begin{equation}
  \label{eq:fhat-bcs}
  \hat{f}_{m (\bvec{k} + \bvec{G})}^{\dagger} = \hat{f}_{m \bvec{k}}^{\dagger}, \qquad
  \hat{f}_{m (\bvec{k} + \bvec{G})} = \hat{f}_{m \bvec{k}}, \qquad \forall \bvec{G} \in \mathcal{R}_{m}^{*}.
\end{equation}
The interacting model for TBG takes the form
\begin{equation}
  \label{eq:tbg-interacting}
  \hat{H} = \hat{H}_{0} - \hat{H}_{\mathrm{sub}} + \hat{H}_{I},
\end{equation}
where $\hat{H}_{0}$ denotes the single-particle Hamiltonian, \(\hat{H}_{\mathrm{sub}}\) denotes the double counting subtraction correction to the single-particle Hamiltonian, and $\hat{H}_{I}$ denotes the electron-electron contribution. 
The choice of double counting subtraction \(\hat{H}_{\mathrm{sub}}\) has a subtle, but important, role in the properties of the ground states \cite{BultinckKhalafLiuEtAl2020,BernevigSongRegnaultEtAl2021,VafekKang2020,XieMacDonald2020,FaulstichStubbsZhuEtAl2023}.
In this work, we consider an interacting model which is projected to the flat bands.
  While the flat bands are separated from the remote bands by a gap, with the inclusion of relaxation, this gap is of a comparable scale to the Coulomb interactions and can alter the resulting many-body ground states.
While these remote bands could be included in principle, due to the additional computational complexity, we leave this to future work.

The single particle Hamiltonian part, \(\hat{H}_{0}\), takes the form
\begin{equation}
  \hat{H}_{0} = \sum_{\bvec{k} \in \mathcal{K}}^{} \sum_{m \in \mathcal{N}}^{} \varepsilon_{m \bvec{k}} \hat{f}_{m \bvec{k}}^{\dagger} \hat{f}_{m \bvec{k}},
\end{equation}
where we recall \(\varepsilon_{m \bvec{k}}\) are the bands of the Bistritzer-MacDonald Hamiltonian \cref{eq:bm_bloch}. The double counting subtraction $\hat{H}_{\text{sub}}$ has a similar form, also quadratic in the creation and annihilation operators; we present this in~\cref{sec:subtr-hamilt} after we have introduced the Coulomb  and exchange  operators.

The electron-electron contribution takes the usual form in terms of the electron repulsion integral (ERI)
\begin{equation}
  \label{eq:h-interacting-original}
  \hat{H}_{I} = \frac{1}{2} \sum_{\bvec{k}, \bvec{k}', \bvec{k}'', \bvec{k}''' \in \mathcal{K}}^{}  \sum_{m,n,m',n' \in \mathcal{N}}^{} \braket{m \bvec{k}, m' \bvec{k}' | n \bvec{k}'', n' \bvec{k}''' } \hat{f}_{m \bvec{k}}^{\dagger} \hat{f}_{m' \bvec{k}'}^{\dagger} \hat{f}_{n' \bvec{k}'''}
  \hat{f}_{n \bvec{k}''},
\end{equation}
where the coefficients are defined as
\begin{equation}
  \braket{m \bvec{k}, m' \bvec{k}' | n \bvec{k}'', n' \bvec{k}''' } = 
  \frac{1}{|\Gamma_{\m}^{\mathcal{K}}|^{2}} \int_{\Gamma_{\m}^{\mathcal{K}}}^{} \int_{\Gamma_{\m}^{\mathcal{K}}} 
  V(\bvec{r} - \bvec{r}') \overline{\psi_{m \bvec{k}}(\bvec{r})} \, \overline{\psi_{m' \bvec{k}'}(\bvec{r}')} \psi_{n \bvec{k}''}(\bvec{r}) \psi_{n' \bvec{k}'''}(\bvec{r}') \mathrm{d}\bvec{r} \mathrm{d}\bvec{r}',
\end{equation}
\(V(\bvec{r})\) denotes the (screened) Coulomb kernel and \(\Gamma_{\m}^{\mathcal{K}}\) denotes a \(n_{k_{x}} \times n_{k_{y}}\) supercell corresponding to the discretization \(\mathcal{K}\), so $|\Gamma_{\m}^{\mathcal{K}}|=n_{k_{x}} n_{k_{y}}|\Gamma_{\m}|$.
We expand the indices $m,m',n,n'$, and use the definition of the eigenfunctions to get
\begin{equation}
 \begin{split} 
  &\braket{(\nu, \eta) \bvec{k}, (\nu', \eta') \bvec{k}' | (\nu'', \eta'') \bvec{k}'', (\nu''', \eta''') \bvec{k}''' } \\
   & \qquad =   \frac{1}{|\Gamma_{\m}^{\mathcal{K}}|^{2}} \int_{\Gamma_{\m}^{\mathcal{K}}}^{} \int_{\Gamma_{\m}^{\mathcal{K}}}  V(\bvec{r} - \bvec{r}')e^{i(\nu''\bvec K_1 - \nu \bvec K_1 + \bvec k''-\bvec k)\cdot \bvec r}e^{i(\nu'''\bvec K_1 - \nu' \bvec K_1 + \bvec k'''-\bvec k')\cdot \bvec r'}  \\
  & \qquad  \qquad   \times   \overline{u_{(\nu, \eta) \bvec{k}}(\bvec{r})} \, \overline{u_{(\nu', \eta') \bvec{k}'}(\bvec{r}')} u_{(\nu'', \eta'') \bvec{k}''}(\bvec{r}) u_{(\nu''', \eta''') \bvec{k}'''}(\bvec{r}') \mathrm{d}\bvec{r} \mathrm{d}\bvec{r}'.
 \end{split}
\end{equation}
The terms $e^{i(\nu''\bvec K_1 - \nu \bvec K_1)\cdot \bvec r}$ and $e^{i(\nu'''\bvec K_1 - \nu' \bvec K_1)\cdot \bvec r'}$ are fast oscillating unless $\nu'' = \nu$ and $\nu''' = \nu'$ because the valleys are far apart. Since $V$ is smooth, the overall integral is thus negligible unless $\nu'' = \nu$ and $\nu''' = \nu'$, and we hence assume that the integrals are non-zero only when $\nu'' = \nu$ and $\nu''' = \nu'$.
After some algebra, we can rewrite \cref{eq:h-interacting-original} in terms of the Fourier coefficients of the pair product corresponding to the same valley:
\begin{equation}
  [\rho_{\bvec{k}, \bvec{k''}}(\bvec{G})]_{(\nu,\eta),(\nu'',\eta'')}
  :=  \delta_{\nu,\nu''}  \frac{1}{|\Gamma_\m|} \int_{{\Gamma}_\m}^{} e^{-i \bvec{G} \cdot \bvec{r}}\overline{u_{(\nu,\eta) \bvec{k}}(\bvec{r})} u_{(\nu'',\eta'') \bvec{k}''}(\bvec{r})  \, \mathrm{d} \bvec{r} .\end{equation}
Recovering  $m,m',n,n'$ to simplify the notation, the interacting Hamiltonian \(\hat{H}_{I}\) can now be written
\begin{equation}
  \label{eq:h-interacting}
  \hat{H}_{I} = \frac{1}{2|\Gamma_{\m}^{\mathcal{K}}|} \sum_{\bvec{k}, \bvec{k}', \bvec{q} \in \mathcal{K}}^{} \sum_{\bvec{G} \in \mathcal{R}_\m^*}^{} \sum_{mnm'n' \in \mathcal{N}}^{} \hat{V}(\bvec{q} + \bvec{G})  [\rho_{\bvec{k}, (\bvec{k} + \bvec{q})}(\bvec{G})]_{mn} [\rho_{\bvec{k}', (\bvec{k}' - \bvec{q})}(-\bvec{G})]_{m'n'} \hat{f}_{m \bvec{k}}^{\dagger} \hat{f}_{m' \bvec{k}'}^{\dagger} \hat{f}_{n' (\bvec{k} - \bvec{q})} \hat{f}_{n (\bvec{k}' + \bvec{q})}.
\end{equation}
In this work, we use the double gate screened potential
\begin{equation}
  \hat{V}(\bvec{q}') = \frac{2\pi}{\epsilon} \frac{\tanh(|\bvec{q}'| d/2)}{|\bvec{q}'|},
\end{equation}
with \(\epsilon = 10.79\) is the relative permittivity and \(d = 30\ \text{nm}\) is the gate distance.

\begin{remark}
  In the literature on twisted bilayer graphene, the expressions for the interacting Hamiltonian \(\hat{H}_{I}\) is typically written in terms of the so-called ``form factor'' \(\Lambda_{\bvec{k}}(\bvec{q}')\) where \(\bvec{k} \in \Gamma_{\m}^{*}\) and \(\bvec{q}' \in \mathbb{R}^{2}\).
  There is a one-to-one correspondence between the form factor and \(\rho_{\bvec{k}, \bvec{k}'}(\bvec{G})\) given by the equation:
  \begin{equation}
    \rho_{\bvec{k}, \bvec{k}'}(\bvec{G}) := \Lambda_{\bvec{k}}(\bvec{k}' - \bvec{k} + \bvec{G}).
  \end{equation}
\end{remark}

\subsection{Hartree-Fock Theory}
\label{sec:hartree-fock-theory}
We recall that a Slater determinant state with \(N_{e}\) electrons can be written as
\begin{equation} \label{eq:HF_state}
  \ket{\Psi_{S}} = \hat{b}_{1}^{\dagger} \cdots \hat{b}_{N_{e}}^{\dagger} \ket{\text{vac}}
\end{equation}
where \(\hat{b}_{i}^{\dagger}\) denotes rotated creation operators
\begin{equation}  \label{eq:Stiefel}
  \hat{b}_{i}^{\dagger} = \sum_{\bvec{k} \in \mathcal{K}}^{} \sum_{n \in \mathcal{N}}^{} \hat{f}_{n \bvec{k}}^{\dagger} [ \Xi(\bvec{k}) ]_{n i}, \quad \text{where} \quad \sum_{\bvec{k} \in \mathcal{K}}^{} \sum_{n \in \mathcal{N}}^{} [ \overline{\Xi(\bvec{k})} ]_{n i}  [ \Xi(\bvec{k}) ]_{n j} = \delta_{i j}, \quad 1 \leq i,j \leq N_e,
\end{equation}
so that $\Xi$ is an $N \times N_e$ matrix, where $N := |\mathcal{N}||\mathcal{K}|$. The set of such matrices \(\Xi\) satisfying the orthonormality condition \eqref{eq:Stiefel} is known as the Stiefel manifold. Hartree-Fock states define a one-body reduced density matrix (1-RDM) by:
\begin{equation}
  [P(\bvec{k}, \bvec{k}')]_{mn} := \braket{\Psi_{S} | \hat{f}_{n \bvec{k}'}^{\dagger} \hat{f}_{m \bvec{k}} | \Psi_{S}} = \sum_{i =1}^{N_{e}} [{\Xi(\bvec{k})}]_{m i} \overline{[\Xi(\bvec{k}')]}_{n i}.
\end{equation}
Clearly, $P \in \mathbb{C}^{N \times N}$ is a rank $N_e$ orthogonal projection. The space of such matrices is known as the Grassmann manifold. We will see that the Hartree-Fock energy depends only on the 1-RDM, so it is natural to optimize over the Grassmann manifold rather than the full Hilbert space of states \eqref{eq:HF_state}. For further discussion of numerical methods leveraging the Stiefel and Grassmann manifold structure see, for example, \cite{vidal2024geometricoptimizationrestrictedopencomplete,doi:10.1137/S0895479895290954}. 




Note that due to the boundary conditions~\cref{eq:fhat-bcs} the 1-RDM is periodic with respect to the moir{\'e} reciprocal lattice:
\begin{equation}
  \label{eq:1-rdm-periodicity}
  P(\bvec{k} + \bvec{G}, \bvec{k}' + \bvec{G}') = P(\bvec{k}, \bvec{k}') \qquad \forall \bvec{G}, \bvec{G}' \in \mathcal{R}_{m}^{*}.
\end{equation}
We will be particularly interested in translation-invariant 1-RDMs, which satisfy the additional condition
\begin{equation} \label{eq:trans_inv}
  P(\bvec{k}, \bvec{k}') =
  \begin{cases}
    P(\bvec{k}) & \bvec{k} = \bvec{k}', \\
    \bvec{0} & \text{otherwise,}
  \end{cases}
\end{equation}
so that the 1-RDM is block-diagonal with respect to relative wave-vector $\bvec{k}$, with $|\mathcal{N}| \times |\mathcal{N}|$ blocks. 

It is important to emphasize that condition \eqref{eq:trans_inv} still allows for non-zero 1-RDM entries coupling wave-vectors $\bvec{k}$ in valley $\bvec{K}$ and $\bvec{k}$ in valley $\bvec{K}'$ and vice versa. Recalling \eqref{eq:eigenfunction_new}, condition \eqref{eq:trans_inv} therefore allows for coupling between the time-reversal pair of total wave-vectors $\tilde{\bvec{K}} + \bvec{k}$ and $- (\tilde{\bvec{K}} + \bvec{k} )$, so that the 1-RDM is not block-diagonal with respect to \emph{total} wave-vector. It can be checked directly that Hartree-Fock states corresponding to 1-RDMs which are block-diagonal with respect to total wave-vector, see \eqref{eq:non_IVC_states}-\eqref{eq:non_IVC_1RDMs} for example, are indeed invariant (up to phase) under translations in each co-ordinate by moir\'e lattice vectors. Hartree-Fock states corresponding to 1-RDMs which are not block-diagonal with respect to total wave-vector, see \eqref{eq:IVC_states}-\eqref{eq:IVC_RDMs} for example, need not be. Intervalley coherent states such as \eqref{eq:IVC_states}-\eqref{eq:IVC_RDMs} also display a factor of $\sqrt{3}$ enlargement of the unit cell at the \emph{atomic} scale because of the interference of wavenumbers near to the distinct Dirac points \cite{BultinckKhalafLiuEtAl2020}. 

States coupling wave-vectors in different valleys while satisfying \eqref{eq:trans_inv} are known as intervalley coherent (IVC).  
IVC states have been theoretically predicted as ground states outside of the chiral limit \cite{BultinckKhalafLiuEtAl2020} and have been observed in experiments \cite{NuckollsLeeOhEtAl2023}. In the chiral limit it is known \cite{StubbsRagoneMacDonaldEtAl2025} that many-body ground states are always translation-invariant in the sense of \eqref{eq:trans_inv}. In particular, ground state 1-RDMs may have non-zero entries between relative wave-vectors $\bvec{k}$ in each valley, but every other intervalley entry must be zero.

We can evaluate energy of the electron-electron interactions of a Hartree-Fock state in terms of the 1-RDM \(P\), the Coulomb (Hartree) operator \(\hat{J}[P]\), and exchange (Fock) operator \(\hat{K}[P]\).
For translation invariant Slater determinants, by Wick's theorem \cite[Section 3.3]{ShavittBartlett2009}, the Coulomb operator and exchange operator can be written as follows:
\begin{align}
  \hat{J}[P] & = \frac{1}{|\Gamma_{m}^{\mathcal{K}}|} \sum_{\bvec{k}, \bvec{k}' \in \mathcal{K}}^{} \sum_{\bvec{G} \in \mathcal{R}^*_\m}^{} \sum_{m, n \in \mathcal{N}}^{}  \hat{V}(\bvec{G}) \tr{\Big( \rho_{\bvec{k}', \bvec{k}'}(-\bvec{G}) P(\bvec{k}') \Big)} [\rho_{\bvec{k},\bvec{k}}(\bvec{G})]_{m n} \hat{f}_{m \bvec{k}}^{\dagger} \hat{f}_{n \bvec{k}}, \label{eq:coulomb-operator} \\
  \hat{K}[P] & = \frac{1}{|\Gamma_{m}^{\mathcal{K}}|} \sum_{\bvec{k}, \bvec{k}' \in \mathcal{K}}^{} \sum_{\bvec{G} \in \mathcal{R}_\m^*}^{} \sum_{m,n,m',n' \in \mathcal{N}}^{} \hat{V}(\bvec{k}' - \bvec{k} + \bvec{G}) [\rho_{\bvec{k}, \bvec{k}'}(\bvec{G})]_{m n} [P(\bvec{k}')]_{n m'} [\rho_{\bvec{k}',
  \bvec{k}}(-\bvec{G})]_{n' m'} \hat{f}_{m \bvec{k}}^{\dagger} \hat{f}_{n' \bvec{k}} \label{eq:exchange-operator}
\end{align}
with corresponding matrix representations $J[P]$ and $K[P]$,
and the electron-electron contribution to the energy can be written:
\begin{equation}
  \braket{\Psi| \hat{H}_{I} | \Psi}
  = \braket{\Psi| \left(\hat{J}[P] - \hat{K}[P] \right) | \Psi }
  = \tr{\Big((J[P] - K[P]) P\Big)}.
\end{equation}

The ground state energy of the Hartree-Fock states can be written as the minimization problem on the matrix $P$
\begin{equation}
    \mathcal E_{\text{HF}} = \min \tr{ \Big( ( H_0 -  H_{\text{sub}} + J[P] - K[P])  P \Big)}, \quad \text{s.t. } P^2=P, \; \tr P = N_e.
\end{equation}
Minimizers can then be computed by self-consistently diagonalizing the Fock matrix $F[P] :=  H_0 -  H_{\text{sub}} + J[P] - K[P]$ and setting $P$ to be the orthogonal projector onto the lowest $N_e$ eigenfunctions (aufbau principle); for more detail, see \cite{szabo1996modern} for example.


\subsection{The Subtraction Hamiltonian}
\label{sec:subtr-hamilt}
As discussed in \cref{sec:bm}, the Bistritzer-MacDonald Hamiltonian is based on a tight binding model for graphene which includes electron-electron effects through a mean-field.
When we directly add the electron-electron effects by adding \(H_{I}\) we therefore ``double count'' electron-electron interactions.
To correct this double counting, a common approach in the literature is to include a double counting subtraction term \cite{BultinckKhalafLiuEtAl2020,BernevigSongRegnaultEtAl2021,VafekKang2020,XieMacDonald2020,FaulstichStubbsZhuEtAl2023}.
While there are a number of different proposals for the double counting subtraction in the literature, a common choice is to use the mean field subtraction which can be written in terms of the Coulomb and exchange operators:
\[
  \hat{H}_{\mathrm{sub}} = \hat{J}[P_{0}] - \hat{K}[P_{0}],
\]
where \(P_{0}\) is a fixed 1-RDM. 
Previous work has shown that the choice of subtraction can modify the nature of many-body ground states \cite{FaulstichStubbsZhuEtAl2023}.
In this work, we choose the ``average'' subtraction, which corresponds to setting \(P_0 \equiv \frac{1}{2} I_{4 \times 4}\) for all \(\bvec{k} \in \mathcal{K}\).
This choice of subtraction gives that the interacting Hamiltonian, $\hat{H}$, is positive semidefinite for the chiral model after an appropriate constant shift \cite{Stubbs_Becker_Lin_2024}.


\subsection{Symmetry Order Parameter}
\label{sec:symm-order-param}
The Bistritzer-MacDonald Hamiltonian (\cref{eq:H_BM}) satisfies a large number of symmetries due to the point group symmetries of the hexagonal lattice as well as the properties of graphene.
While these symmetries are initially defined on the single particle level, they play an critical role in understanding the nature of the many-body ground states \cite{BultinckKhalafLiuEtAl2020,BernevigSongRegnaultEtAl2021,BeckerLinStubbs2023,SongBernevig2022}.

Following \cite{FaulstichStubbsZhuEtAl2023}, we define a family of order parameters for the symmetries \(C_{2z} \mathcal{T}\), \(\nu_{x} \mathcal{T}\), and \(\nu_{y} \mathcal{T}\).
To show how these order parameters are defined, let us consider \(C_{2z} \mathcal{T}\); the other other two symmetries follow by a similar argument.
For full details, we refer to \cite[Appendices D, E]{FaulstichStubbsZhuEtAl2023}.

The symmetry \(C_{2z} \mathcal{T}\) is an antiunitary symmetry for which \((C_{2z}\mathcal{T}) \Hbm (\bvec{k}) = \Hbm (\bvec{k}) (C_{2z} \mathcal{T})\).
This transformation is characterized by the \textit{sewing matrix} \(B_{\bvec{k}}(C_{2z} \mathcal{T})\) which is an \( |\mathcal{N}| \times  |\mathcal{N}|\) matrix defined as follows
\begin{equation}
  \label{eq:sewing_matrix}
  \begin{split}
    [B_{\bvec{k}}(C_{2z} \mathcal{T})]_{mn}
    & = \braket{ u_{m \bvec{k}}, (C_{2z} \mathcal{T})u_{n \bvec{k}}} \\
    & = \sum_{\bvec{G} \in \mathcal{R}_{\m}^{*}}^{} \sum_{\sigma, \ell}^{} \overline{u_{m \bvec{k}}(\bvec{G}; \sigma, \ell)}\, \overline{u_{n \bvec{k}}(\bvec{G}; -\sigma, \ell)}.
  \end{split}
\end{equation}
Using the fact that \(C_{2z} \mathcal{T}\) commutes with \(H(\bvec{k})\) it's easy to see that \(B_{\bvec{k}}(C_{2z} \mathcal{T})\) is a unitary matrix.

The sewing matrix, \(B_{\bvec{k}}(C_{2z} \mathcal{T})\), encodes the action of \(C_{2z} \mathcal{T}\) on a basis of eigenvectors at each \(\bvec{k}\).
Since the 1-RDM is defined in terms of these eigenvectors, we can use the sewing matrix to detect whether a certain symmetry is satisfied or not.
For \(C_{2z} \mathcal{T}\), the symmetry transforms the 1-RDM as
\begin{equation}
  P(\bvec{k}) \mapsto B_{\bvec{k}}(C_{2z} \mathcal{T})\, \overline{P(\bvec{k})} \, B_{\bvec{k}}(C_{2z} \mathcal{T})^{-1}.
\end{equation}
Hence to determine symmetry breaking, we can consider the order parameter:
\begin{equation}
  O_{C_{2z} \mathcal{T}} = \frac{1}{N_{\mathcal{K}}} \sum_{\bvec{k} \in \mathcal{K}}^{} \| B_{\bvec{k}}(C_{2z} \mathcal{T})\, \overline{P(\bvec{k})} \, B_{\bvec{k}}(C_{2z} \mathcal{T})^{-1} - P(\bvec{k}) \|,
\end{equation}
where $N_{\mathcal{K}}=n_{k_{x}} n_{k_{y}}$, which measures the deviation of the 1-RDM from the value expected if the symmetry were satisfied.

For \(\nu_{x} \mathcal{T}\), the sewing matrix \(B_{\bvec{k}}(\nu_{x} \mathcal{T})\) is defined analogously to~\cref{eq:sewing_matrix}.
In this case, the corresponding order parameter can be written
\begin{equation}
  O_{\nu_{x} \mathcal{T}} = \frac{1}{N_{\mathcal{K}}} \sum_{\bvec{k} \in \mathcal{K}}^{} \| B_{\bvec{k}}(\nu_{x} \mathcal{T})\, \overline{P(-\bvec{k})} \, B_{\bvec{k}}(\nu_{x} \mathcal{T})^{-1} - P(\bvec{k}) \|,
\end{equation}
where the additional minus sign is due to the fact that \(\nu_{x} \mathcal{T}\) satisfies \((\nu_{x} \mathcal{T}) H(\bvec{k}) = H(-\bvec{k}) (\nu_{x} \mathcal{T})\).
The symmetry order parameter for \(\nu_{y} \mathcal{T}\) has an identical formula to \(\nu_{x} \mathcal{T}\) with the sewing matrix \(B_{\bvec{k}}(\nu_{x} \mathcal{T})\) replaced with \(B_{\bvec{k}}(\nu_{y} \mathcal{T})\).

\subsection{Gauge Fixing and Initialization}\label{sec:gauge-fixing-init}
Due to the existence of many local minima, proper initialization play a critical role in ensuring the Hartree-Fock SCF iterations converge to true minimizer of the Hartree-Fock energy.
To find proper initialization, we begin by recounting the analysis of the Bistritzer-MacDonald model with \(w_{0} = 0\) which is called the ``chiral limit''.

At the chiral limit, the IBM model is exactly solvable by Hartree-Fock theory; that is, a many-body state if a ground state if and only if it can be written as a linear combination of Hartree-Fock states which are also ground states \cite{StubbsRagoneMacDonaldEtAl2025}.
Previous studies have shown that the many-body ground states away from the chiral limit can be well approximated by the ground states found at the chiral limit \cite{FaulstichStubbsZhuEtAl2023,SoejimaParkerBultinckEtAl2020,ChatterjeeBultinckZaletel2020,KwanWagnerBultinck2021,BultinckKhalafLiuEtAl2020}.

One of the key steps underlying the analysis at the chiral limit is that there exists a family of unitaries \(\{ U(\bvec{k}) : \bvec{k} \in \mathcal{K} \} \subseteq U(|\mathcal{N}|)\) so that after the basis transformation
\begin{equation}
  \hat{f}_{m \bvec{k}}^{\dagger} \mapsto \sum_{n}^{} \hat{f}_{n \bvec{k}}^{\dagger} [U(\bvec{k})]_{n m}, \qquad 
  \hat{f}_{m \bvec{k}} \mapsto \sum_{n}^{} \hat{f}_{n \bvec{k}} [\overline{U(\bvec{k})}]_{n m} ,
\end{equation}
the pair product \(\rho_{\bvec{k}, \bvec{k}'}(\bvec{G})\) is a diagonal matrix for all \(\bvec{k}, \bvec{k}' \in \mathcal{K}\) and all \(\bvec{G} \in \mathcal{R}_{m}^{*}\).
This choice of basis is referred to as the \textit{sublattice polarized gauge}, since in this basis the Bloch eigenfunctions \(\psi_{m \bvec{k}}\) are all fully supported on either the \(A\) or \(B\) sublattice.
Therefore, we can write the creation operators as \(\hat{f}_{(\bvec{K}, A), \bvec{k}}^{\dagger}, \hat{f}_{(\bvec{K}', A), \bvec{k}}^{\dagger}, \hat{f}_{(\bvec{K}, B),\bvec{k}}^{\dagger}, \text{or } \hat{f}_{(\bvec{K}', B), \bvec{k}}^{\dagger}\) where \(\hat{f}_{(\nu, \sigma), \bvec{k}}^{\dagger}\) creates a state at momentum \(\bvec{k}\) in valley \(\nu\) on sublattice \(\sigma\).
Away from the chiral limit, it is not possible to find a basis choice so \(\psi_{m \bvec{k}}\) are fully sublattice polarized; however we can still search for a basis which is maximally localized on each of the sublattices.
For this purpose, we use the \(k\)-resolved selected columns of the density matrix (SCDM-\(k\)) to localize on each of the sublattices \cite{DamleLinYing2015,DamleLinYing2017}.

Once we have found the sublattice polarized gauge using SCDM-\(k\), we will initialize our calculations using five candidate ground states which are exact ground states at the chiral limit.
 The quantum hall (QH), valley hall (VH), valley polarized (VP), Kramers intervalley coherent (KIVC), and time reversal intervalley coherent (TIVC) states.
These states are called ``ferromagnetic'' Slater determinant states because their 1-RDM \(P(\bvec{k})\) is independent of \(\bvec{k}\). They can be split into non-intervalley coherent states and intervalley coherent states. 

The non-intervalley coherent states are given explicitly by 
\begin{equation} \label{eq:non_IVC_states}
  \ket{\text{QH}} := \prod_{\bvec k \in \mathcal K} \hat{f}^\dagger_{(\bvec{K} ,\mathrm A) \bvec k} \hat{f}^\dagger_{(\bvec{K}',\mathrm B) \bvec k} \ket{\text{vac}}, \quad \ket{\text{VH}} := \prod_{\bvec k \in \mathcal K} \hat{f}^\dagger_{(\bvec{K},\mathrm A) \bvec k} \hat{f}^\dagger_{(\bvec{K}',\mathrm A) \bvec k} \ket{\text{vac}}, \quad \ket{\text{VP}} := \prod_{\bvec k \in \mathcal K} \hat{f}^\dagger_{(\bvec{K}, \mathrm A) \bvec k} \hat{f}^\dagger_{(\bvec{K},\mathrm B) \bvec k} \ket{\text{vac}}.
\end{equation}
Since which electrons states are occupied at each \(\bvec{k}\) is the same, these states can be equivalently expressed by their 
1-RDM
\begin{equation} \label{eq:non_IVC_1RDMs}
  P_{\mathrm{QH}} =
  \begin{bmatrix}
    1 & & & \\
      & 0 & & \\
      & & 0 & \\
      & & & 1
  \end{bmatrix},
  \qquad
  P_{\mathrm{VH}} =
  \begin{bmatrix}
    1 & & & \\
      & 0 & & \\
      & & 1 & \\
      & & & 0
  \end{bmatrix},
  \qquad
  P_{\mathrm{VP}} =
  \begin{bmatrix}
    1 & & & \\
      & 1 & & \\
      & & 0 & \\
      & & & 0
  \end{bmatrix}.
\end{equation}
The intervalley coherent states are families of states that can be  parameterized by $\varphi$ as follows:
\begin{equation} \label{eq:IVC_states}
\begin{split}
    \ket{\text{KIVC}} := \prod_{\bvec k \in \mathcal K} \frac{1}{2}\left(ie^{-i\varphi}\hat{f}^\dagger_{(\bvec{K}, \mathrm A) \bvec k} - \hat{f}^\dagger_{(\bvec{K'}, \mathrm B) \bvec k} \right) \left(ie^{-i\varphi} \hat{f}^\dagger_{(\bvec{K},\mathrm B) \bvec k} + \hat{f}^\dagger_{(\bvec{K'}, \mathrm A) \bvec k} \right) \ket{\text{vac}}, \\
     \ket{\text{TIVC}} := \prod_{\bvec k \in \mathcal K} \frac{1}{2}\left(e^{-i\varphi}\hat{f}^\dagger_{(\bvec{K}, \mathrm A) \bvec k} + \hat{f}^\dagger_{(\bvec{K'}, \mathrm B) \bvec k} \right) \left(e^{-i\varphi} \hat{f}^\dagger_{(\bvec{K},\mathrm B) \bvec k}  + \hat{f}^\dagger_{(\bvec{K'}, \mathrm A) \bvec k} \right) \ket{\text{vac}},
    \end{split}
\end{equation}
and are equivalently expressed by
\begin{equation} \label{eq:IVC_RDMs}
  P_{\mathrm{KIVC}} =
  \frac{1}{2}
  \begin{bmatrix}
    1 & & & -ie^{-i \varphi} \\
    & 1 & ie^{-i \varphi} & \\
    & -ie^{i \varphi} & 1 & \\
    ie^{i \varphi} & & & 1
  \end{bmatrix},
 \qquad 
  P_{\mathrm{TIVC}} =
  \frac{1}{2}
  \begin{bmatrix}
    1 & & & e^{-i \varphi} \\
    & 1 & e^{-i \varphi} & \\
    & e^{i \varphi} & 1 & \\
    e^{i \varphi} & & & 1
  \end{bmatrix}.
\end{equation}
These states can be differentiated in a gauge independent way using the order parameters introduced in~\cref{sec:symm-order-param}; see \cref{tab:ground_state_symmetry}.
\begin{table}[h!]
  \centering    
  \renewcommand{\arraystretch}{1.4}
  \begin{tabular}{cccc}
    \toprule
    & $O_{C_{2z} \mathcal{T}}$ & $O_{\nu_{x} \mathcal{T}}$ & $O_{\nu_{y} \mathcal{T}}$ \\ \midrule
    QH & 1 & 1 & 1 \\
    VH & 1 & 0 & 0 \\
    VP  & 0 & 1 & 1 \\
    KIVC & \(|\Re{(e^{i \varphi})}|\) & 1 & 0 \\
    TIVC & \(|\Im{(e^{i \varphi})}|\) & 0 & 1 \\ \bottomrule
  \end{tabular}
  \caption{The different ground states and their expected symmetries. Note that the $C_{2z} \mathcal{T}$ order parameter of KIVC and TIVC states depend on the choice of $\varphi$.}
  \label{tab:ground_state_symmetry}
\end{table}
  
We note that these 1-RDMs are only used as \textit{initialization}; to find the Hartree-Fock ground states we must first perform an SCF calculation.
Typically, the Hartree-Fock ground states are close to the idealized 1-RDMs which can be checked by the matching of the order parameters.

\section{Application: Simulation of Interacting BM model with 
Structural Relaxation} \label{sec:numerics}

\subsection{Numerical model and experiments}

In this work we perform calculations at the Hartree-Fock (HF) and Coupled Cluster Singles and Doubles (CCSD) levels of theory; for a quick review of both methods as applied to TBG we refer to \cite[Sections IV.A and IV.C]{FaulstichStubbsZhuEtAl2023}
All calculations presented in this section are performed using Python-based Simulations of Chemistry Framework (PySCF) \cite{Sun_Berkelbach_Blunt_Booth_Guo_Li_Liu_McClain_Sayfutyarova_Sharma_etal_2018,Sun_Zhang_Banerjee_Bao_Barbry_Blunt_Bogdanov_Booth_Chen_Cui_etal_2020}. We precompute the one- and two-electron integrals for the Hamiltonian.
PySCF allows us to perform HF and CCSD calculations, and evaluate energetics and 1-RDM for symmetry detection. In this paper, we perform a spinless calculation. We use a Morkhorst-Pack grid $\mathcal K$ with $n_{k_x} = n_{k_y} = 6$. We study the system at charge neutrality, which corresponds to the total number of electrons being $N_e = \sum_{\bvec k} \tr P(\bvec k) = 2 n_{k_x} n_{k_y} = 72$. At each moir\'e cell there are two electrons and a total of four orbitals. 

The interacting model for TBG contains multiple states that are energetically close to the ground state. Without careful initialization, it is often difficult to converge to the global minimum in energy. 
To ensure we converge to a ground state, we consider adiabatically evolving a single particle Hamiltonian from the chiral model, where the interacting model is exactly solvable, to a target Hamiltonian.
In particular, we consider the following two families of Hamiltonians which interpolate between the chiral limit and a target Hamiltonian:
\begin{equation} \label{eq:interp}
\begin{split}
  H_0^{\mathrm{BM}}(\kappa) & = (1-\kappa) \Hchi + \kappa \HbmVV, \quad 0 \leq \kappa \leq .95, \\
  H_{0}^{\mathrm{relaxed}}(\alpha) & =   (1-\alpha) \Hchi + \alpha \HrelVV, \quad 0 \leq \alpha \leq 1.
  \end{split}
\end{equation}
For each choice of \(\alpha\) and \(\kappa\), we construct the corresponding interacting model and initialize with the five states introduced in \cref{sec:gauge-fixing-init}.
We observe that in each case, the HOMO-LUMO gap for each initialization and each value of \(\alpha\) and \(\kappa\), does not close suggesting the interpolation in~\cref{eq:interp} is adiabatic.

It is important to stress that \(H_{0}^{\mathrm{BM}}(\kappa)\) and \(H_{0}^{\mathrm{relaxed}}(\alpha)\) use different methods for modeling relaxation and therefore cannot be directly compared.
For a rough comparison, we can estimate the value of $\kappa$ corresponding to our relaxed model by calculating the ratio between the AA and AB terms in the first shell of \(\HrelVV\). We find that $[\tilde T_1]_{\A\A} / [\tilde T_1]_{\A\B} \approx 0.7$ (see \cref{sec:relax-bm-inter}) so we expect that \(H_{0}^{\mathrm{BM}}(\kappa = 0.7) \approx \HrelVV\).
We refer to $\kappa=0.7$ as the ``physical ratio'' of the BM model. This agrees with previous works on the effects of relaxation in effective models, where the ratio is estimated to be between 0.7 and 0.8 \cite{Nam_Koshino_2017,Fang_Carr_Zhu_Massatt_Kaxiras_2019,Kang_Vafek_2023,Vafek_Kang_2023}.

\subsection{Energy differences}
\begin{figure}[h!]
\centering
\begin{subfigure}{.3\textwidth}
\includegraphics[width=\columnwidth]{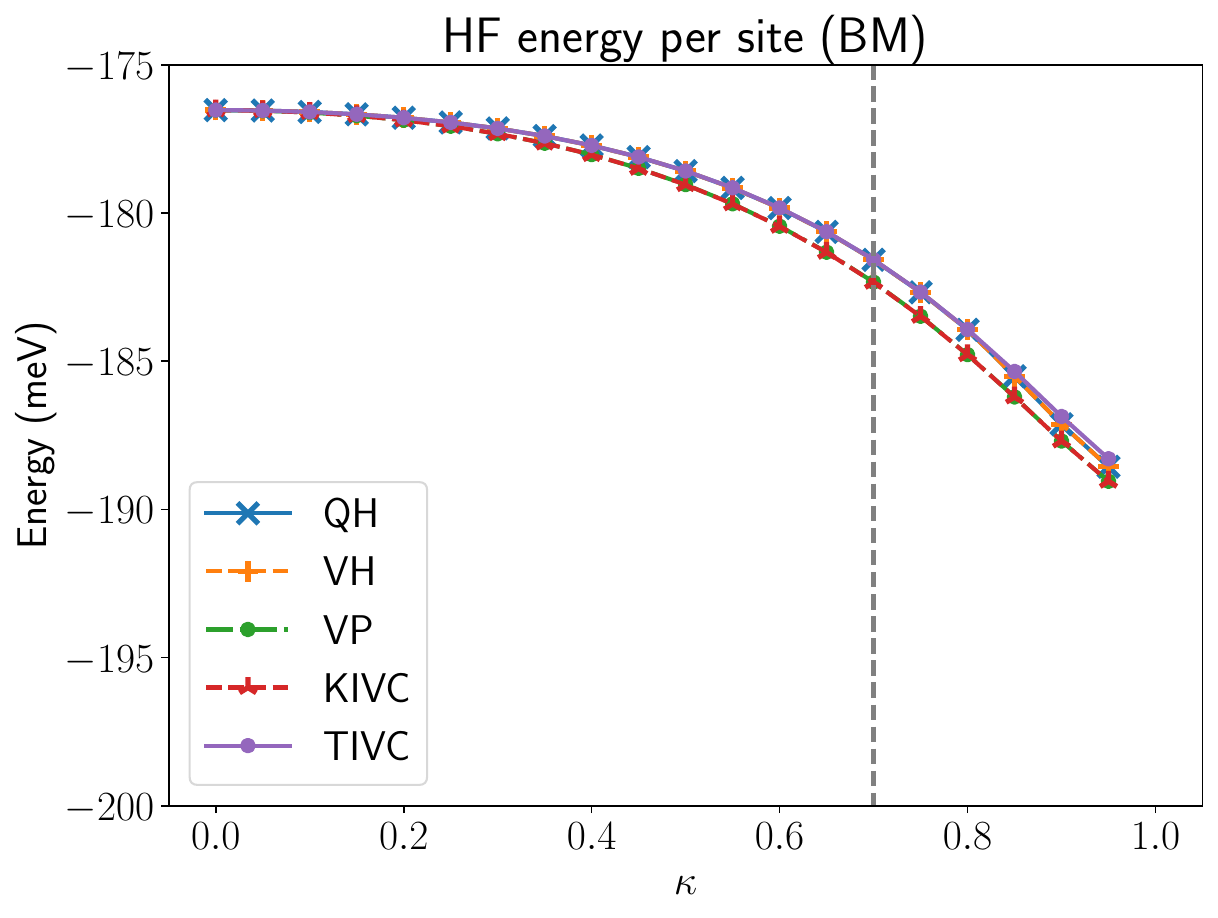}
\caption{\label{fig:bm_e}}
\end{subfigure}
\begin{subfigure}{.3\textwidth}
\includegraphics[width=\columnwidth]{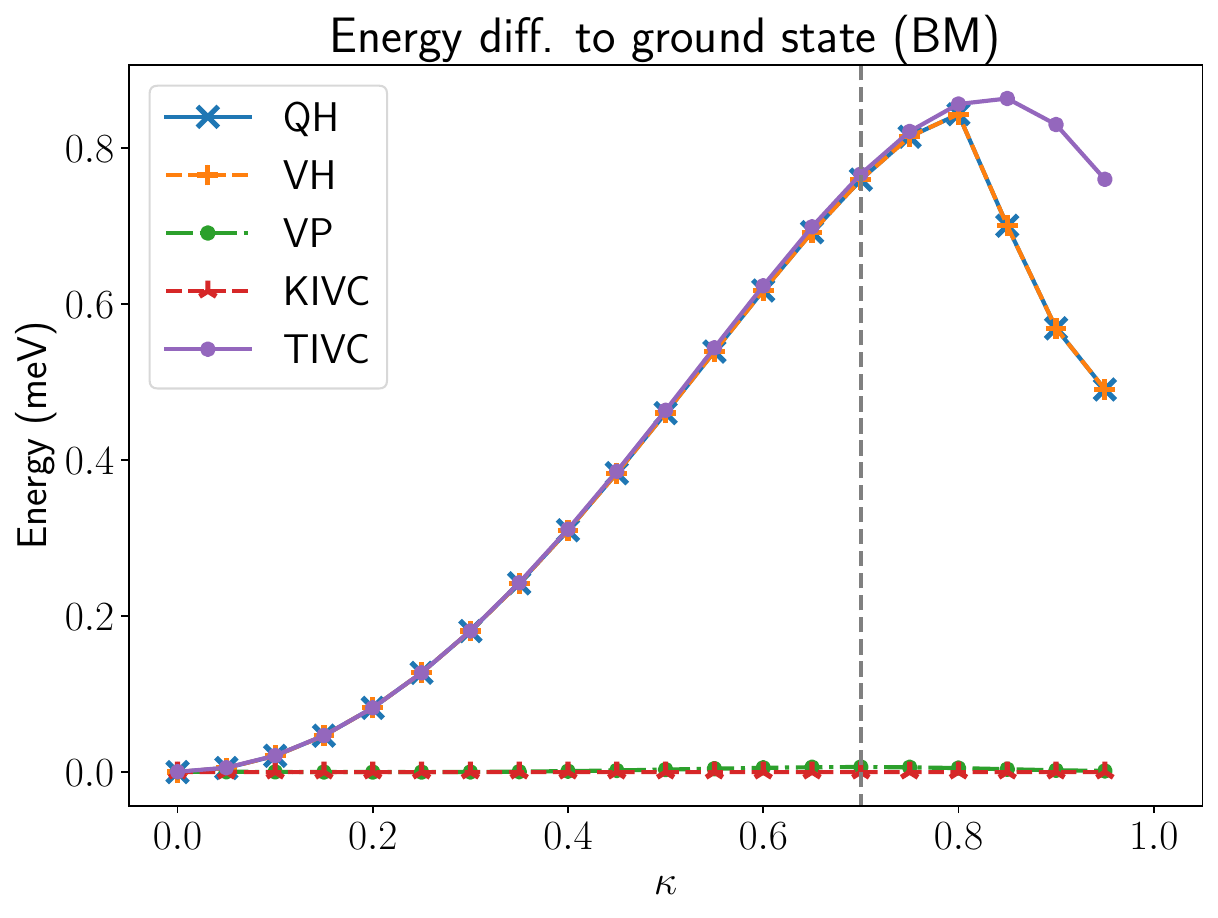}
\caption{\label{fig:bm_e_diff}}
\end{subfigure}
\begin{subfigure}{.3\textwidth}
\includegraphics[width=\columnwidth]{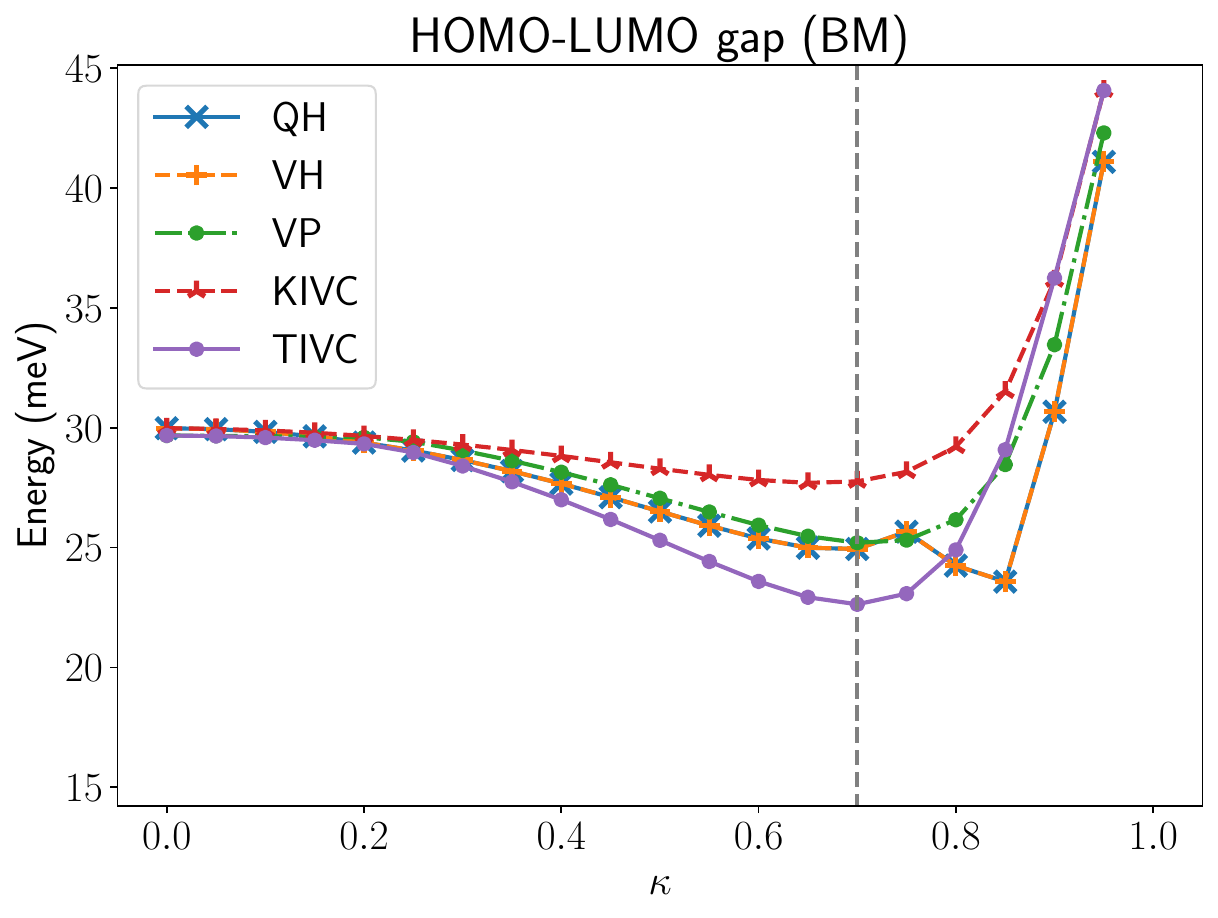}
\caption{\label{fig:bm_gap}}
\end{subfigure}\\[.2in]
\begin{subfigure}{.3\textwidth}
\includegraphics[width=\columnwidth]{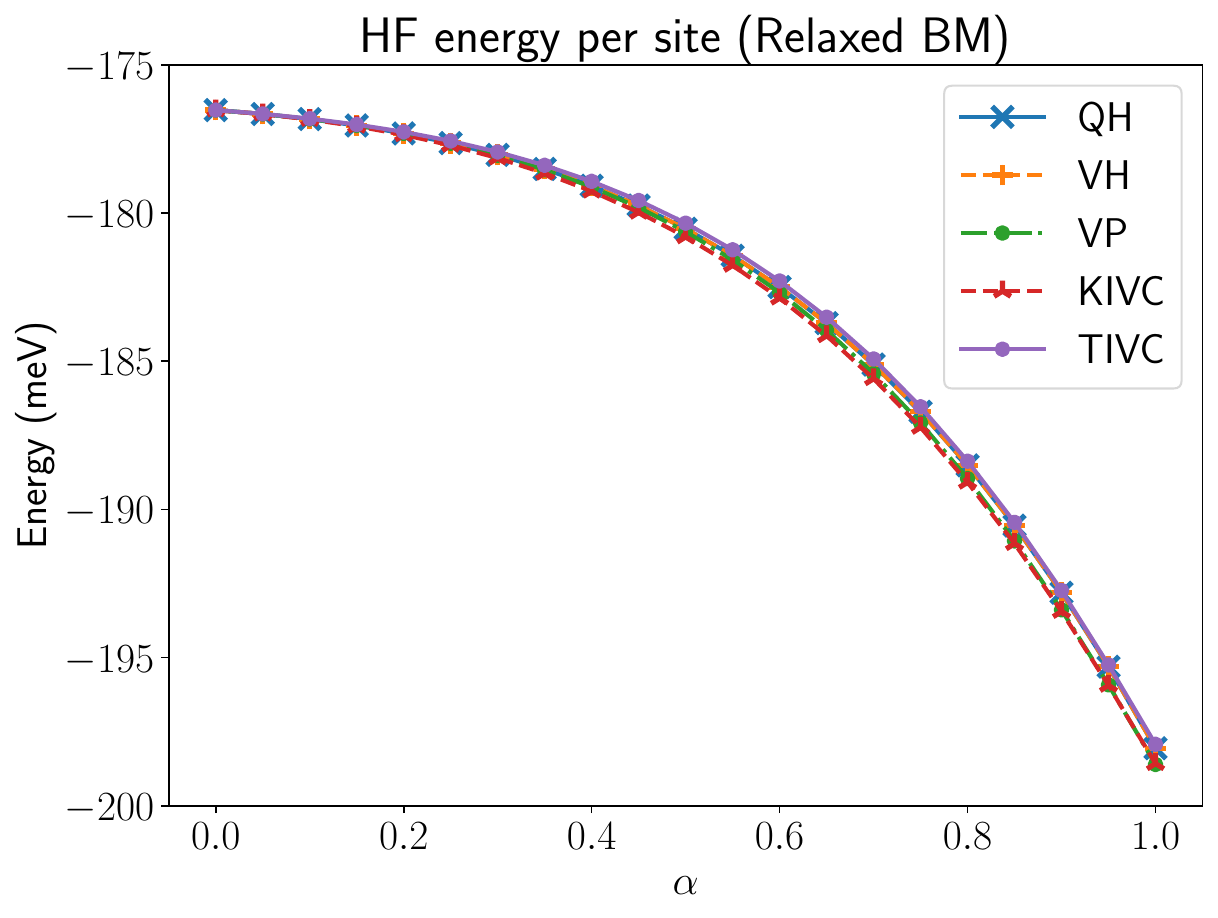}
\caption{\label{fig:relax_e}}
\end{subfigure}
\begin{subfigure}{.3\textwidth}
\includegraphics[width=\columnwidth]{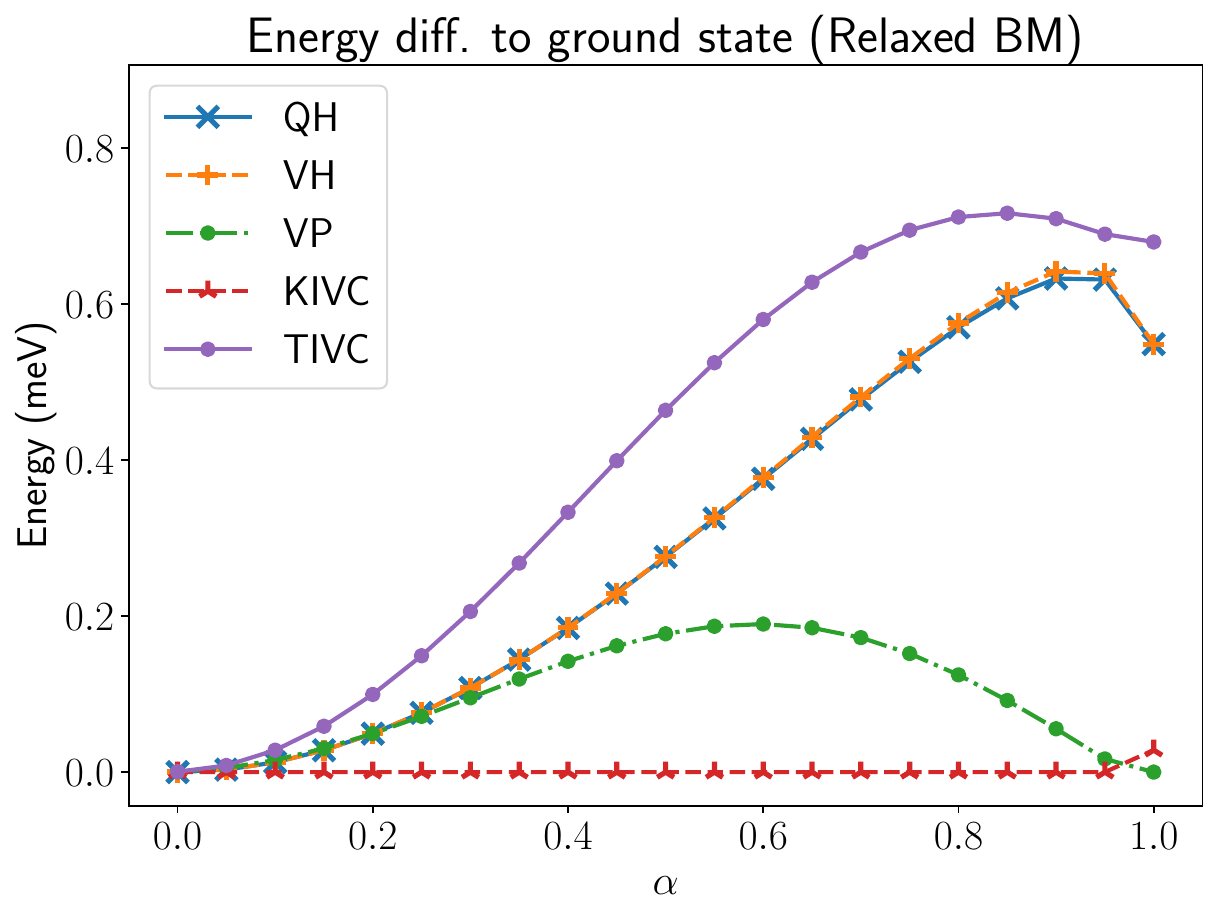}
\caption{\label{fig:relax_e_diff}}
\end{subfigure}
\begin{subfigure}{.3\textwidth}
\includegraphics[width=\columnwidth]{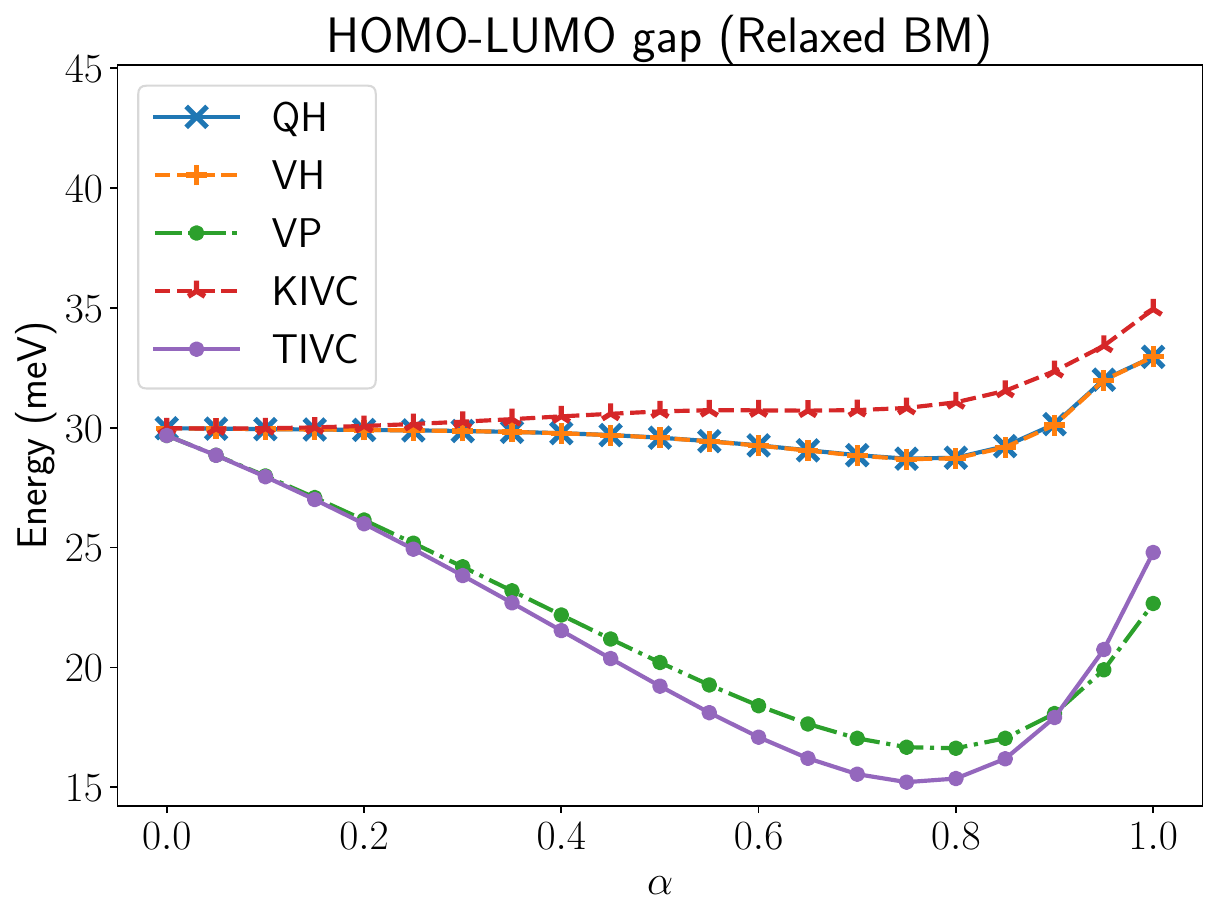}
\caption{\label{fig:relax_gap}}
\end{subfigure}
\caption{Energy of converged state per moir\'e site, energy difference to ground state (minimum of the five candidate states), and HOMO-LUMO gap with respect to interpolation parameter $\alpha$ for the BM model  (\subref{fig:bm_e}, \subref{fig:bm_e_diff}, \subref{fig:bm_gap}) and for the relaxed BM model (\subref{fig:relax_e}, \subref{fig:relax_e_diff}, \subref{fig:relax_gap}). The gray line ($\kappa=0.7$) represents the physical ratio of relaxation in the BM model.}
\end{figure}

The energy of converged HF states decrease as a function of $\kappa$ and $\alpha$ in both models, see \cref{fig:bm_e,fig:relax_e}. Different initializations converge to different energies, suggesting the energy landscape of this problem is complicated. The energy variation per moir\'e site is less than 1 meV across initializations. In the BM model, both KIVC and VP are degenerate ground states for $\kappa \in [0, 0.95]$, see \cref{fig:bm_e_diff}. In the relaxed BM model, KIVC state is the ground state for $\alpha \in [0, 0.95]$, while VP state is the ground state for $\alpha = 1$, see \cref{fig:relax_e_diff}. 
 We can also establish the energy ordering of states as $\alpha$ varies. At the chiral limit ($\kappa = \alpha = 0$), our numerical result is consistent with the theory that all five states are degenerate ground states. The  BM model has the energy ordering KIVC $\approx$ VP < QH $\approx$ VH $\approx$ TIVC  when $\kappa \leq 0.8$, and KIVC $\approx$ VP < QH $\approx$ VH < TIVC for $\kappa > 0.8$. The additional splitting of QH and VH from TIVC can be attributed to a phase transition, also observed in the symmetry order parameters (see \cref{fig:bm_c2zt}). The relaxed BM model has the energy ordering KIVC < VP < QH $\approx$ VH < TIVC for values of $\alpha$ below 0.95 and at $\alpha = 1$, the VP state has lower energy than KIVC by 0.02 meV. The HOMO-LUMO gap remains large for all interpolation coefficient and initialization in both models, indicating that the system is an insulator at charge neutrality. 

\subsection{Symmetry breaking}
We also investigate the effects of interpolation on the symmetry order parameters. In both models, the symmetry order parameters of VP, KIVC and TIVC states stay relatively constant, indicating that these states does not undergo phase transition (see \cref{fig:symm}). Based on converged 1-RDM and symmetry order parameters, we conclude that these three states remain candidate ground states away from the chiral limit. In contrast, states with QH and VH initializations undergo a phase transition. \cref{fig:bm_c2zt} show a transition from a $C_{2z}\mathcal T$ broken phase to a $C_{2z}\mathcal T$ symmetric phase at around $\kappa = 0.8$ for the BM model, and \cref{fig:relax_c2zt} shows a similar transition at around $\alpha = 0.95$ for relaxed BM. These transitions at the same value of $\kappa$ and $\alpha$ are also observed in QH initialization for $\nu_x \mathcal T$ and $\nu_y \mathcal T$ symmetries (see \cref{fig:bm_tpx,fig:relax_tpx,fig:bm_tpy,fig:relax_tpy}). The phase transition changes QH/VH states into a state characterized by
\begin{equation}
  P(\bvec k)= \frac{1}{2} 
  \begin{bmatrix}
    1 & -1 & & \\
    -1 & 1 & & \\
      & & 1 & -1\\
      & & -1& 1
  \end{bmatrix}.
 \end{equation}
The transition of the $C_{2z}\mathcal T$ symmetry order parameter is also observed in the single valley interacting BM model \cite{FaulstichStubbsZhuEtAl2023}. This is expected, as the valleyful QH/VH states can be viewed as two copies of single valley QH states, one in each valley. 

\subsection{Differences between models}
\begin{figure}[h!]
\centering
\begin{subfigure}{.3\textwidth}
\includegraphics[width=\columnwidth]{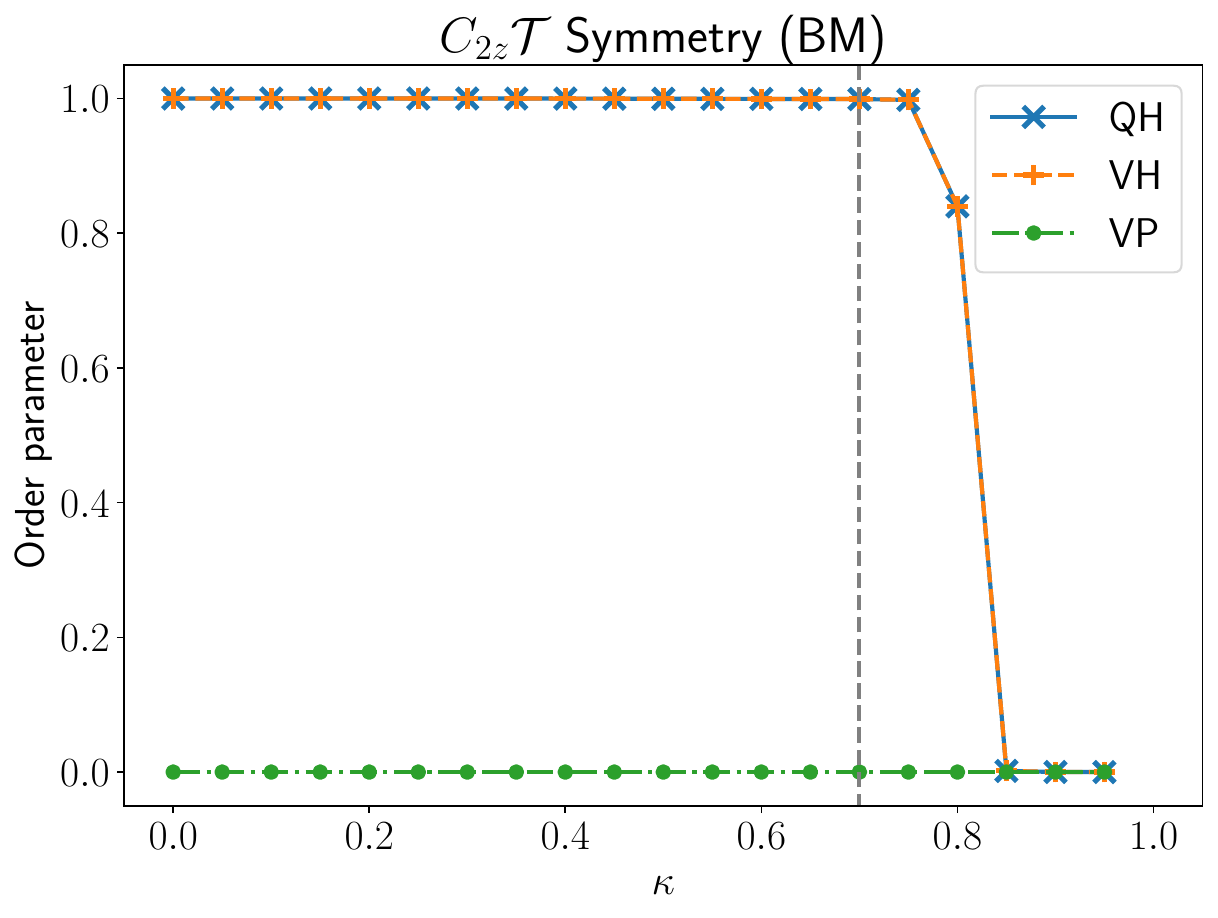}
\caption{\label{fig:bm_c2zt}}
\end{subfigure}
\begin{subfigure}{.3\textwidth}
\includegraphics[width=\columnwidth]{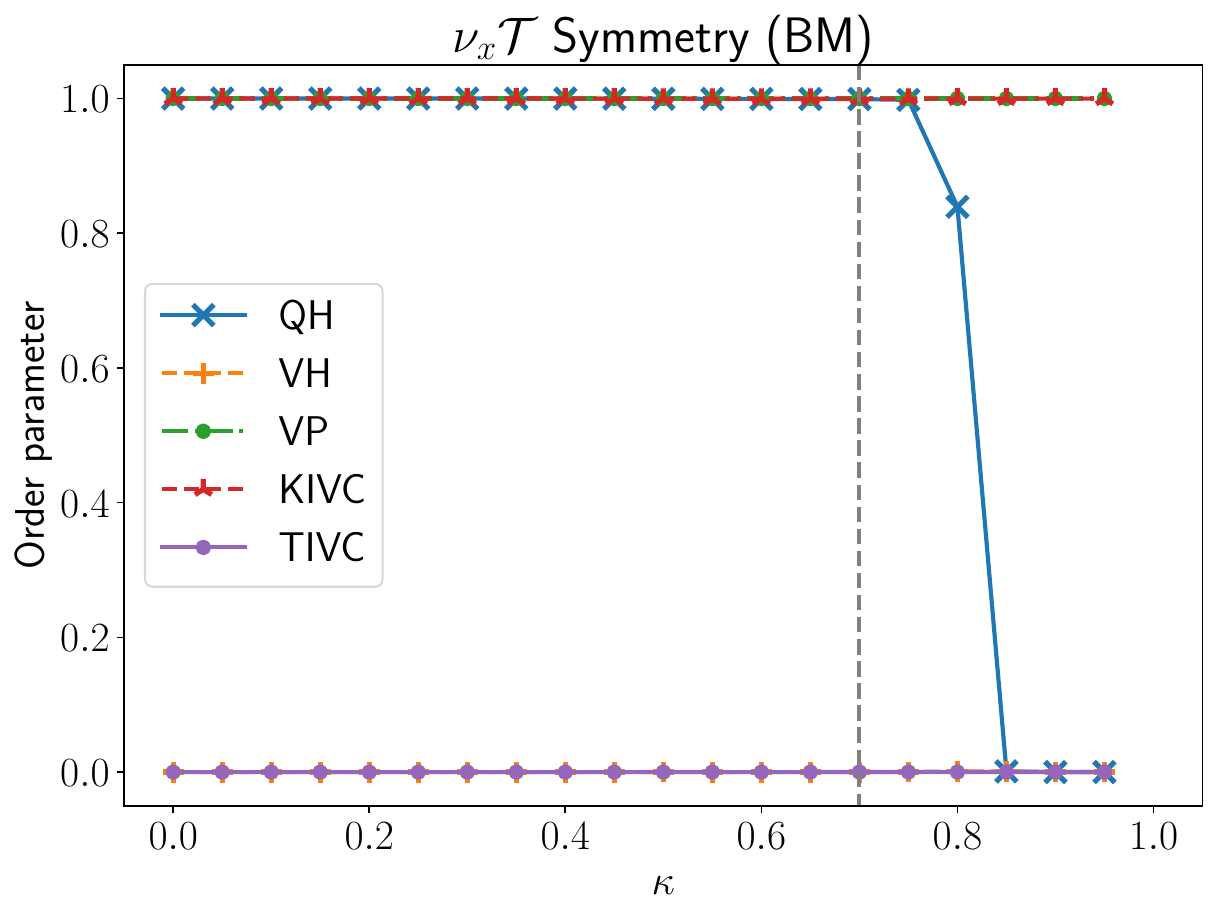}
\caption{\label{fig:bm_tpx}}
\end{subfigure}
\begin{subfigure}{.3\textwidth}
\includegraphics[width=\columnwidth]{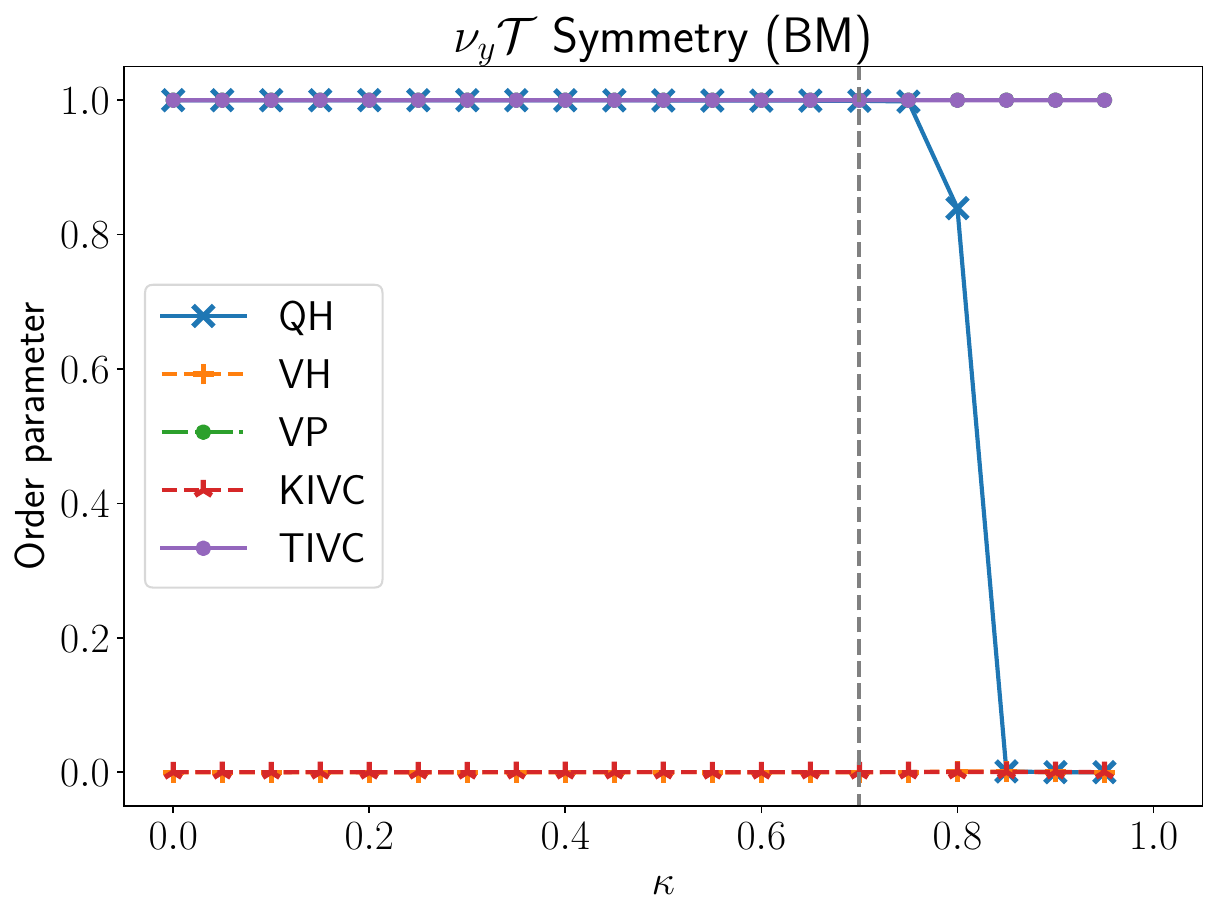}
\caption{\label{fig:bm_tpy}}
\end{subfigure}\\[.2in]
\begin{subfigure}{.3\textwidth}
\includegraphics[width=\columnwidth]{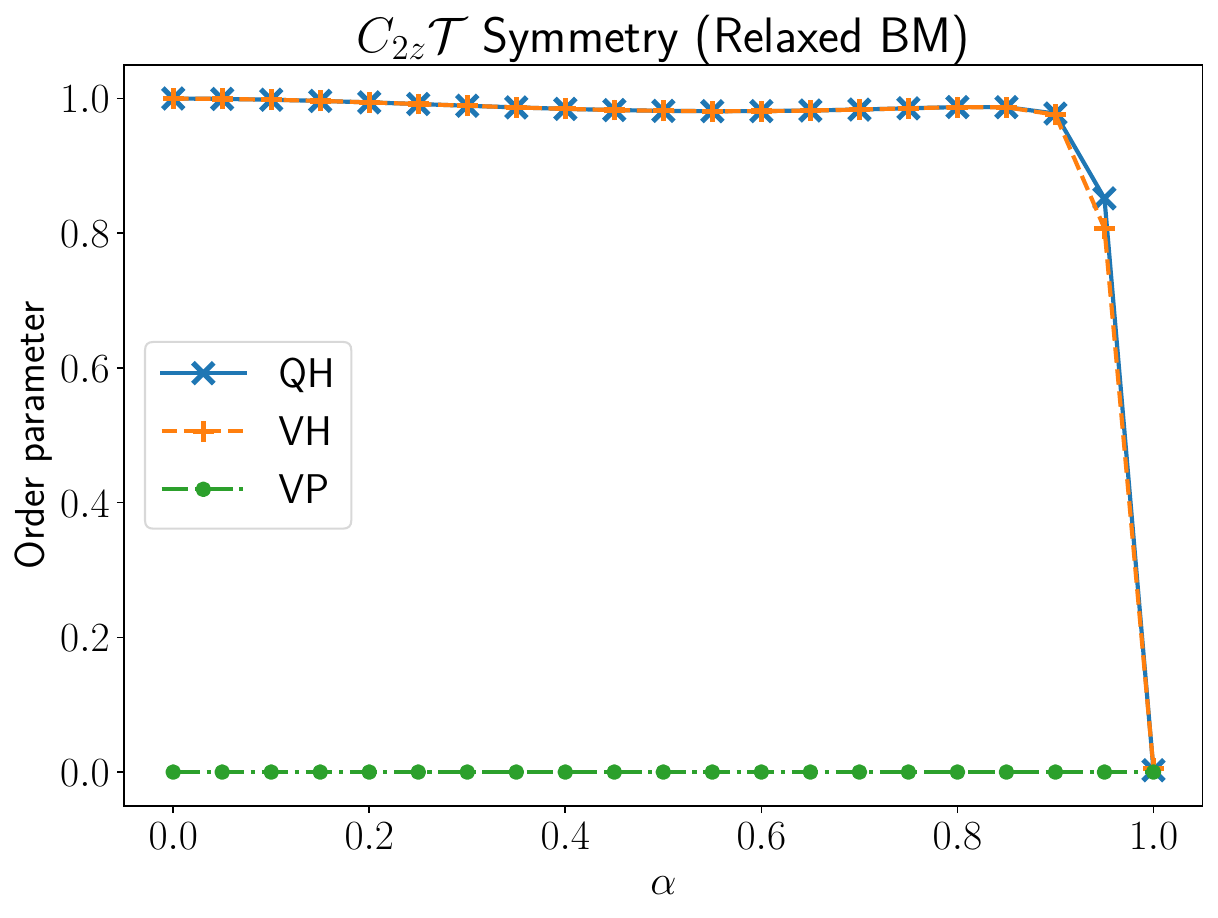}
\caption{\label{fig:relax_c2zt}}
\end{subfigure}
\begin{subfigure}{.3\textwidth}
\includegraphics[width=\columnwidth]{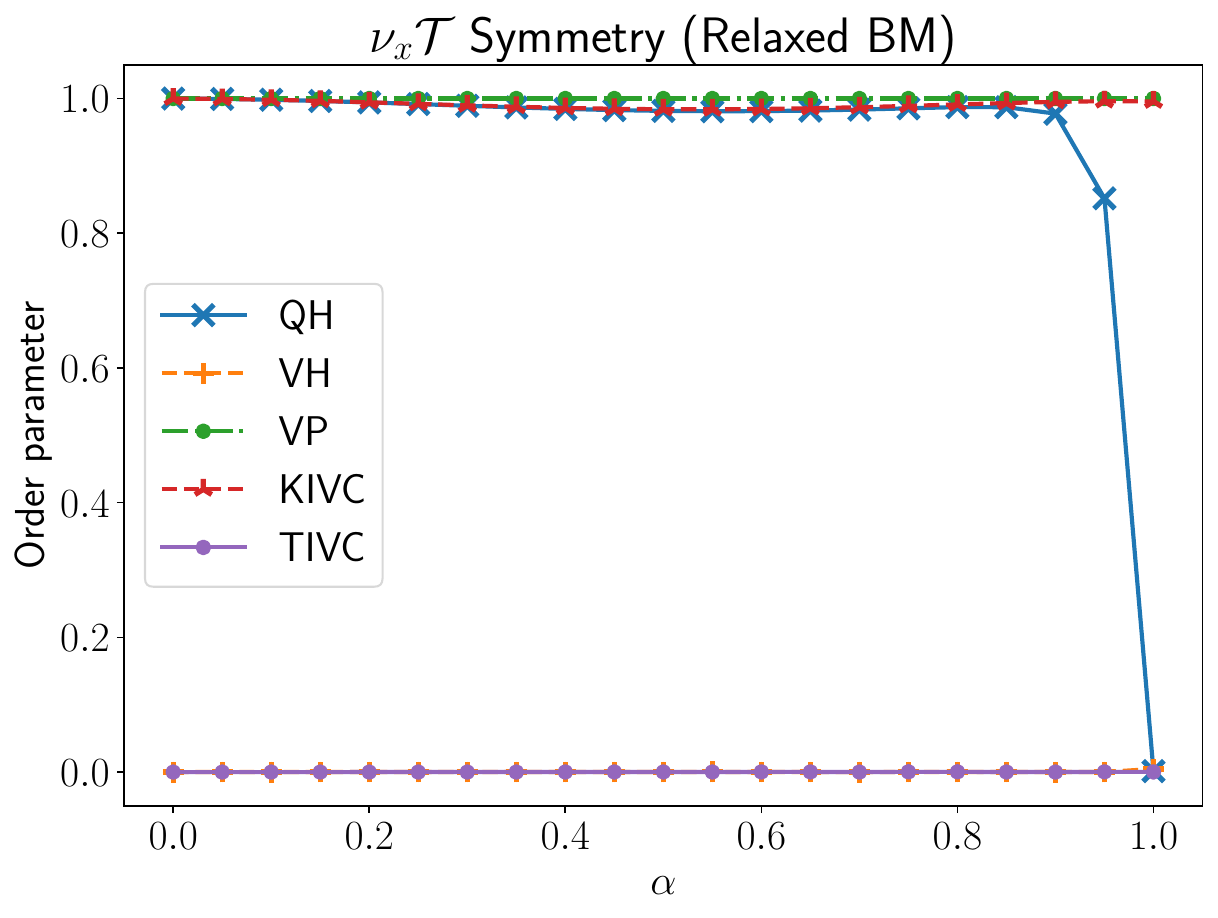}
\caption{\label{fig:relax_tpx}}
\end{subfigure}
\begin{subfigure}{.3\textwidth}
\includegraphics[width=\columnwidth]{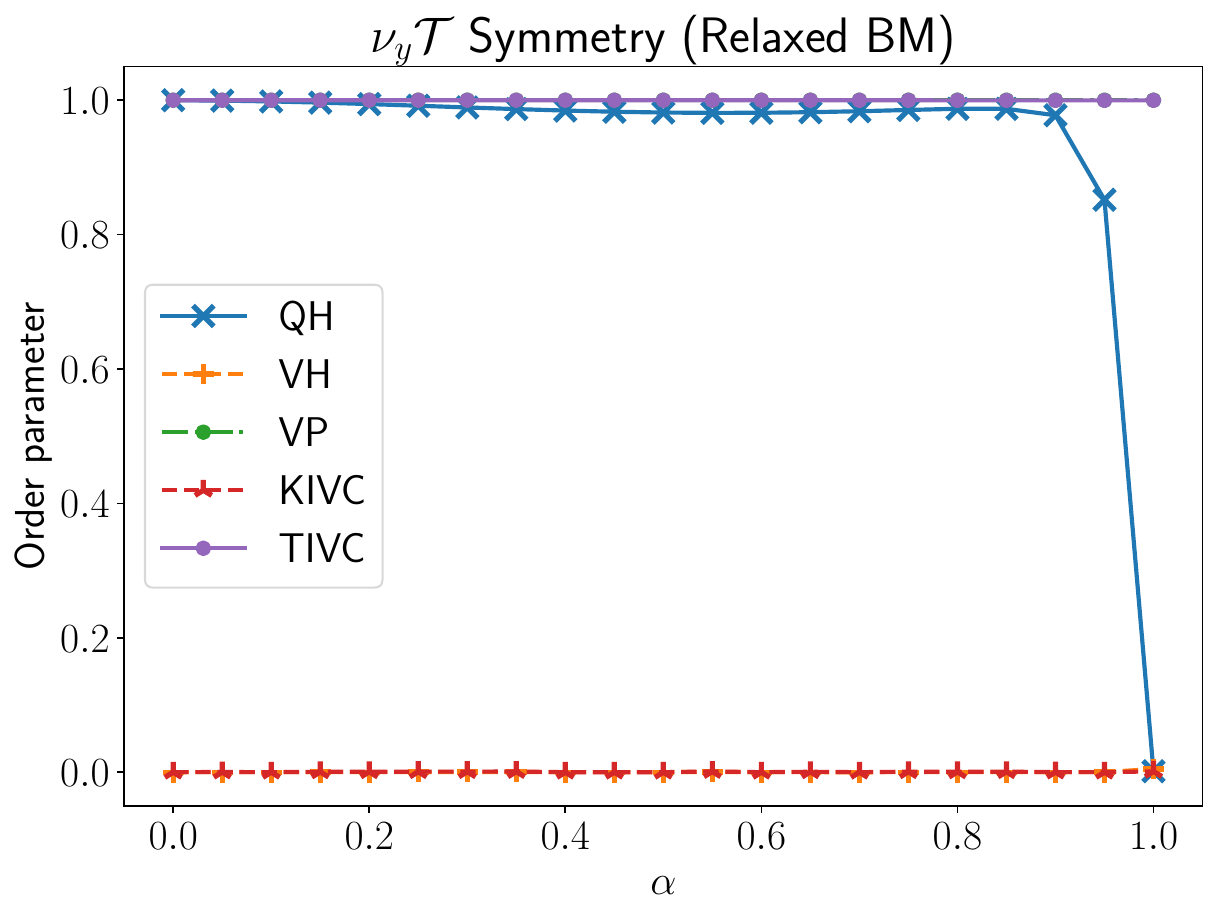}
\caption{\label{fig:relax_tpy}}
\end{subfigure}
\caption{Symmetry order parameters for BM (\subref{fig:bm_c2zt}, \subref{fig:bm_tpx}, \subref{fig:bm_tpy}) and relaxed BM model (\subref{fig:relax_c2zt}, \subref{fig:relax_tpx}, \subref{fig:relax_tpy}). QH, VH and VP initializations are plotted for $C_{2z}\mathcal T$ symmetry, and all five initializations are plotted for $\nu_x\mathcal T$ and $\nu_y\mathcal T$ symmetries. The gray line ($\kappa=0.7$) represents the physical ratio of relaxation in the BM model. \label{fig:symm}}
\end{figure}

We comment on the discrepancies between the relaxed BM model and the ordinary BM model with relaxation accounted for by tuning the ratio of AA to AB hopping. When considering both models in the relaxed configuration (BM model at $\kappa = 0.70$ and relaxed BM model at $\alpha = 1$), we observe the BM model predicts a degenerate ground state of KIVC and VP state, while the relaxed BM model favors the VP state slightly (see \cref{fig:bm_e_diff,fig:relax_e_diff}). Furthermore, the BM model predicts $C_{2z}\mathcal T$ broken QH and VH states, while relaxed BM predicts $C_{2z}\mathcal T$ symmetric states (see \cref{fig:bm_c2zt,fig:relax_c2zt}). Similar discrepancy also occurs for $\nu_x \mathcal T$ and $\nu_y \mathcal T$ symmetries. In fact, the symmetries of the relaxed BM model at $\alpha = 1$ are comparable to those of the BM model with ratio $\kappa \approx 0.85$. We attribute such underestimation of the AA and AB hopping ratio to the inclusion of intralayer scattering, longer range interlayer hopping and non-local momentum terms in the relaxed BM model. It is interesting that symmetry-breaking phase transitions seem to occur so close to where $\alpha = 1$, i.e., exactly at the structure predicted by our relaxation model. We will investigate this phenomenon in more detail in future work.

\subsection{Correlation energy}
\begin{figure}[h!]
\centering
\begin{subfigure}{.3\textwidth}
\includegraphics[width=\columnwidth]{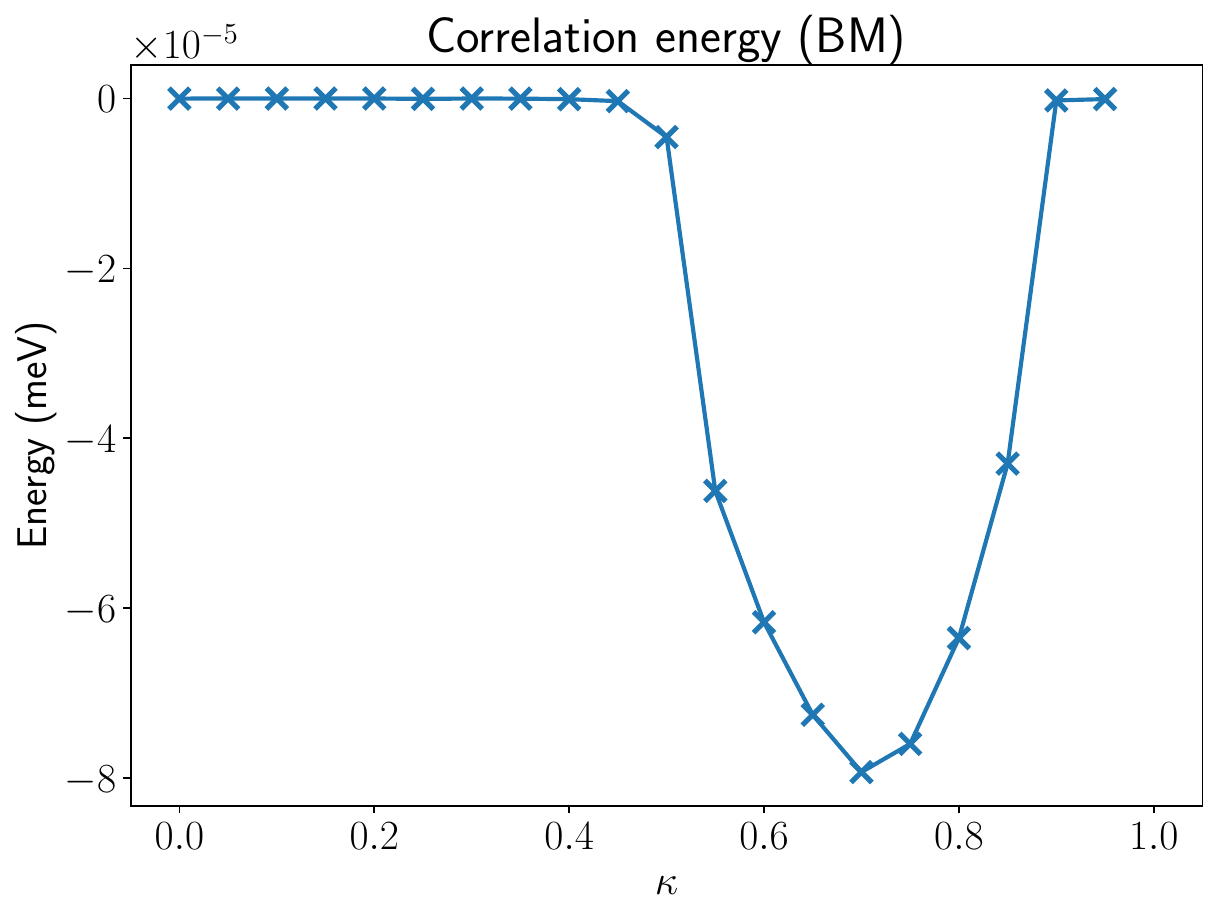}
\caption{\label{fig:bm_e_corr}}
\end{subfigure}
\begin{subfigure}{.3\textwidth}
\includegraphics[width=\columnwidth]{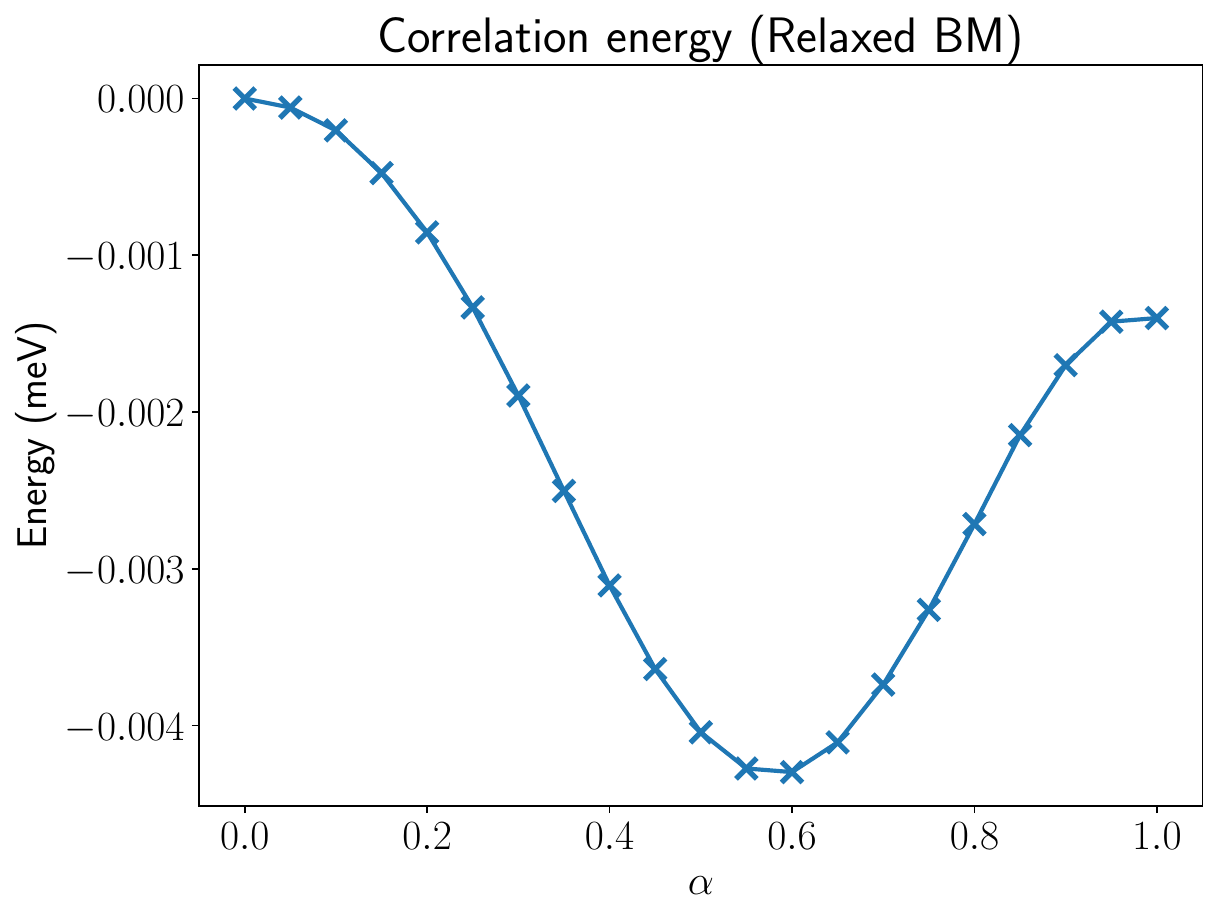}
\caption{\label{fig:relax_e_corr}}
\end{subfigure}
\caption{Correlation energy per moir\'e site for the BM model \subref{fig:bm_e_corr} and the relaxed BM \subref{fig:relax_e_corr} model.}
\end{figure}

Finally, we present the correlation energy, defined as the difference between CCSD and HF ground state energies. All energies are reported per moir\'e site. \cref{fig:bm_e_corr,fig:relax_e_corr} shows that the correlation energy is extremely small ($\sim 10^{-5} \text{ meV}$ for BM and $\sim 10^{-3} \text{ meV}$ for relaxed BM) compared to the energy scale of HF ground state ($\sim 100 \text{ meV}$). At the chiral limit, the correlation energy vanishes, consistent with the  theoretical prediction that HF states are the exact ground states. Surprisingly, the correlation energy also vanishes for the BM model with $\kappa < 0.4$, suggesting the theoretical prediction may hold under small perturbations of $\kappa$. We also observe the correlation energy reaches a minimum at around $\kappa = 0.7$ for the BM model, and $\alpha = 0.6$ for the relaxed BM model.

\section{Conclusion} \label{sec:conclusion}

In this work we have introduced a novel many-body model of twisted bilayer graphene's electronic properties which systematically accounts for the effects of structural relaxation. We have also computed minimizers of this model within the Hartree-Fock and CCSD levels of theory. While we find that the relative ordering of candidate ground states is fairly consistent with simpler treatments of relaxation, we also find that our model predicts the significantly earlier onset of symmetry-breaking phase transitions. In particular, our model predicts that these transitions should occur right at the structure predicted by our relaxation model. We will investigate this phenomenon in more detail in future work. The framework introduced in the present paper can also be adapted to include the effects of strain. Strain is known to introduce new candidate many-body ground states \cite{NuckollsLeeOhEtAl2023,KwanWagnerBultinck2021}, making this an exciting future direction.

\section*{Acknowledgement}
TK's, ML's, and KS's  research was partially supported by Simons Targeted Grant Award No. 896630. ML's research was also partially supported in part by NSF Award No. 2422469, and  AW's research was supported in part by grant NSF DMS-2406981.

ML's research was also supported in part by grant NSF PHY-2309135 to the Kavli Institute for Theoretical Physics (KITP) during a visit to KITP, and ML also thank the Flatiron Institute for hospitality while (a portion of) this research was carried out. The Flatiron Institute is a division of the Simons Foundation.  TK thanks the DOE Lawrence Berkeley Laboratory for supporting a summer internship to work on this project.

The authors acknowledge helpful conversations with Lin Lin of UC Berkeley/DOE Lawrence Berkeley Laboratory and Chao Yang of DOE Lawrence Berkeley Laboratory.

\bibliographystyle{plain}
\bibliography{bibliography.bib}

\appendix

\section{Detailed Description of Relaxed BM model}
We utilize the symmetry of configuration space to help deriving the effective relaxed BM model. We introduce the local configuration based relaxation
\begin{equation}
	\bvec u_1(\bvec x) = \bvec U_{\m, 1}(\vec\gamma_2^{-1} (\bvec x)) , 
	\quad \bvec u_2(\bvec x) = \bvec U_{\m, 2}(\vec\gamma_1^{-1} (\bvec x)),
\end{equation}
and such displacements are periodic with respect to the lattice on the opposite layer
\begin{equation}
	\bvec u_1(\bvec x + \bvec R_2) = \bvec u_1(\bvec x), \quad \bvec u_2(\bvec x + \bvec R_1) = \bvec u_2(\bvec x), \quad \bvec R_\ell \in \mathcal R_\ell .
\end{equation}
With these newly defined displacements, we can rewrite \cref{eq:displacement} as
\begin{equation}
	\bvec R_1 + \vec \tau_1^\sigma \mapsto \bvec R_1 + \vec \tau_1^\sigma + \bvec u_1(\bvec R_1 + \vec \tau_1^\sigma - \vec \tau_2^\sigma), \quad 	\bvec R_2 + \vec \tau_2^\sigma \mapsto \bvec R_2 + \vec \tau_2^\sigma + \bvec u_2(\bvec R_2 + \vec\tau_2^\sigma - \vec \tau_1^\sigma).
\end{equation}

\subsection{Intralayer terms} \label{sec:relax-bm-intra}
The strategy for deriving the continuum model is substituting the wave packet ansatz \cref{eq:ansatz} into the relaxed tight-binding model \cref{eq:H_tb_relax}.

The intralayer components of \cref{eq:H_tb_relax} can be simplified into
\begin{equation}
	\sum_{\bvec R_1'}\sum_{\sigma'}h_{11, \bvec u}^{\sigma\sigma'}(\bvec R_1 + \vec \tau_1^\sigma - \bvec R_1' - \vec \tau_1^{\sigma'}; \bvec b_1^\sigma)\Psi_{\bvec R_1'}^{\sigma'},
	\end{equation}
with a disregistry dependent hopping function
\begin{equation}
 h_{11, \bvec u}^{\sigma\sigma'}(\bvec Q; \bvec b) =  h_{11}(\bvec Q + \bvec u_1(\bvec b) - \bvec u_1(\bvec b - \bvec Q + \vec\tau_2^\sigma - \vec \tau_2^{\sigma'})),
\end{equation}
and the local disregistry
\begin{equation}
	\bvec b_1^\sigma = \bvec R_1 + \vec \tau_1^\sigma - \vec \tau_2^\sigma \mod \mathcal R_2.
\end{equation}

We can then define the Bloch transform of ${h_{11, \bvec u}^{\sigma\sigma'}}(\bvec Q; \bvec b)$ with both arguments. Notice that the first argument takes in lattice vectors of layer 1 and the second argument is periodic with respect to the lattice vectors of layer 2. The Bloch transformed function is then 
\begin{equation}
	\widetilde{h_{11, \bvec u}^{\sigma\sigma'}}(\bvec k; \bvec G_2) = \sum_{\bvec R_1 \in \mathcal R_1} e^{-i\bvec k \cdot (\bvec R_1 + \vec\tau_1^{\sigma} - \vec\tau_1^{\sigma'})} \frac{1}{|\Gamma_2|}\int_{\Gamma_2} e^{-i\bvec G_2\cdot\bvec b} h_{11, \bvec u}^{\sigma\sigma'}((\bvec R_1 + \vec\tau_1^{\sigma} - \vec\tau_1^{\sigma'}); \bvec b) \dee \bvec b,
\end{equation}
and the effective intralayer Hamiltonian is
\begin{equation}
\begin{split}
	\left[ H_{11}\psi \right] (\bvec r; \sigma, 1) =  & \left[ D^{(1)}(-i\nabla) \cdot \psi\right] (\vec r; \sigma, 1)\\
	& \quad + \sum_{\sigma'} \sum_{j = 1}^{12}  e^{-i \bvec P_j  \cdot \bvec r} \left[\widetilde {h_{11,\bvec u}^{\sigma\sigma'}}(\bvec K_1; \mathcal G_2(\bvec P_j) ) +  \nabla_{\bvec k} \widetilde {h_{11,\bvec u}^{\sigma\sigma'}}(\bvec K_1; \mathcal G_2(\bvec P_j)) \cdot (-i\nabla)\right] \psi(\bvec r; \sigma', 1),
\end{split}
\end{equation}
where $D^{(\ell)}(\bvec k) = v [\sigma_0 - i \sin(\theta_\ell)\sigma_3] \cdot \vec \sigma \cdot \bvec k + v_1 \sigma_0 \bvec k^2 + v_2 [\sigma_1 (-k_1^2 + k_2^2) + 2\sigma_2  k_1 k_2] $ is the Dirac operator with second order and linearized rotation corrections, and the values are listed in \cref{tab:dirac}. The momentum shift vectors $\bvec P_j$ are moir\'e reciprocal lattice vectors (see blue points in \cref{fig:mom_hops}), and $\mathcal G_\ell (\bvec G) = \rot{\ell}\left[\rot 1 - \rot 2 \right]^{-1} \bvec G$ maps moir\'e reciprocal lattice vectors to corresponding reciprocal lattice vectors on layer $\ell$.  

After a Bloch-Floquet transform, the intralayer hopping matrices in \cref{eq:relaxed_hopping} are thus given by
\begin{equation}
	\left[A_j^{(\ell)}\right]_{\sigma\sigma'} = \widetilde {h_{\ell\ell,\bvec u}^{\sigma\sigma'}}(\bvec K_\ell; \mathcal G_{3-\ell}(\bvec P_j) ), 
	\quad \left[A_{j,\nabla}^{(\ell)}\right]_{\sigma\sigma'} = \nabla_{\bvec k}\widetilde {h_{\ell\ell,\bvec u}^{\sigma\sigma'}}(\bvec K_\ell; \mathcal G_{3-\ell}(\bvec P_j)).
\end{equation}
We report some values of these matrices in \cref{tab:intra_params}.
\begin{table}[h!]
\centering
  \renewcommand{\arraystretch}{1.3}
\begin{tabular}{c c c}
\toprule
 $v$ & $v_1$ & $v_2$ \\ \midrule
    5.339 eV$\cdot$\r{A} & -0.783 eV$\cdot \text{\r{A}}^2$   & -3.405 eV$\cdot \text{\r{A}}^2$     \\
 \bottomrule
\end{tabular} 
\caption{The Dirac cone and its second order correction parameters.}\label{tab:dirac}
\end{table}
\begin{table}[h!]
\centering
  \renewcommand{\arraystretch}{1.3}
\begin{tabular}{c c c c c}
\toprule
 $j$ & $\left[A_j^{(1)}\right]_{\A\A}$ (meV) & $\left[A_j^{(1)}\right]_{\A\B}$ (meV) & $\left[A_{j,\nabla}^{(1)}\right]_{\A\A}$ (meV$\cdot$\r{A}) &  $\left[A_{j,\nabla}^{(1)}\right]_{\A\B}$ (meV$\cdot$\r{A}) \\ 
 \midrule
    1 & -0.02 & $14.72-8.13i$   &  $(20.41, 11.04 )$   & $(16.48+8.99i,-9.65+16.51i)$  \\
    4 & 0.02 & $15.61-8.65i$ & $(20.77, 11.45)$ & $(16.68+9.76i, -9.09+16.64i)$ \\
    7 & -0.02 & $0.26+0.18i$ & $(0.36, -0.23)$  & $(0.28-0.18i, 0.18+0.31i)$   \\ 
    10 & 0.02 & $0.28+0.15i$ & $(0.36, -0.22)$  & $(0.32-0.17i, 0.17+0.28i)$   \\ 
 \bottomrule
\end{tabular} 
\caption{Selected parameters for intralayer relaxation parameters for the first two shells of layer 1 for TBG at $1.05^\circ$.  Within the same shell, momentum hoppings related by $2\pi/3$ rotation (for example indexed by 1,3,5 or 2,4,6 in \cref{fig:mom_hops}) have the same magnitude, and only differ by a relative phase.} \label{tab:intra_params}
\end{table}

\subsection{Interlayer terms}\label{sec:relax-bm-inter}
The interlayer component of  \cref{eq:H_tb_relax} can be simplified into
\begin{equation}
	\sum_{\bvec R_2}\sum_{\sigma'}h_{12,\bvec u}^{\sigma\sigma'}(\bvec R_1 + \vec \tau_1^\sigma - \bvec R_2 - \vec\tau_2^{\sigma'})\Psi_{\bvec R_2}^{\sigma'},
\end{equation}
where the relaxation dependent hopping function is given by
\begin{equation}
	h_{12,\bvec u}^{\sigma\sigma'}(\bvec x) := h_{12}(\bvec x  +  \bvec u_1(\bvec x -  \vec\tau_2^\sigma + \vec \tau_2^{\sigma'})  - \bvec u_2( - \bvec x + \vec \tau_1^{\sigma} -\vec\tau_1^{\sigma'})).
\end{equation}
We then define its Fourier transform
\begin{equation}
\widehat{h_{12,\bvec u}^{\sigma\sigma'}}(\bvec k) := \int_{\mathbb R_2} e^{-i\bvec k \cdot \bvec x} h_{12, \bvec u}^{\sigma\sigma'}(\bvec x) \dee \bvec x.
\end{equation}

We identify the leading order effective interlayer Hamiltonian as
\begin{equation}
\begin{split}
   \left[ H_{12}\psi \right] (\bvec r; \sigma, 1) = &\sum_{\sigma'} \sum_{j = 1}^{12} e^{-i \bvec Q_j  \cdot \bvec r} e^{i\mathcal G_2(\vec Q_j) \cdot (\vec \tau_2^\sigma - \vec \tau_2^{\sigma'})}  \\
    & \qquad \times
    \frac{1}{|\Gamma_2|} \left[
     \widehat {h_{12,\bvec u}^{\sigma\sigma'}}(\bvec K_2 + \mathcal G_2(\bvec Q_j) ) +  \nabla \widehat {h_{12,\bvec u}^{\sigma\sigma'}}(\bvec K_2+ \mathcal G_2(\bvec Q_j)) \cdot (-i\nabla)\right] \psi(\bvec r; \sigma', 2).
\end{split}
\end{equation}
The interlayer hopping matrices in \cref{eq:relaxed_hopping} are thus given by
\begin{equation}
	\left[\tilde T_j \right]_{\sigma\sigma'} =  \frac{1}{|\Gamma_2|}e^{i\mathcal G_2(\vec Q_j) \cdot (\vec \tau_2^\sigma - \vec \tau_2^{\sigma'})} 
     \widehat {h_{12,\bvec u}^{\sigma\sigma'}}(\bvec K_2 + \mathcal G_2(\bvec Q_j)), 
	\quad \left[\tilde T_{j,\nabla}\right]_{\sigma\sigma'} = \frac{1}{|\Gamma_2|}e^{i\mathcal G_2(\vec Q_j) \cdot (\vec \tau_2^\sigma - \vec \tau_2^{\sigma'})} \nabla \widehat {h_{12,\bvec u}^{\sigma\sigma'}}(\bvec K_2 + \mathcal G_2(\bvec Q_j)).
\end{equation}

\begin{table}[h!]
  \centering
  \renewcommand{\arraystretch}{1.3}
\begin{tabular}{c c c c c}
\toprule
 $j$ & $\left[\tilde T_j\right]_{\A\A}$ (meV) & $\left[\tilde T_j\right]_{\A\B}$ (meV) & $\left[ \tilde T_{j,\nabla}\right]_{\A\A}$ (meV$\cdot$\r{A})  &  $\left[\tilde T_{j,\nabla}\right]_{\A\B}$ (meV$\cdot$\r{A}) \\ 
 \midrule
    1 & $78.58-2.21i$ & $113.25-3.09i$     & $(-87.2 +0.58i, 0.85+0.05i)$   & $(-93.54+1.45i, 0.87-80.40i)$   \\
    4 & $-0.36+0.02i$   & $-1.57+3.01i$     & $(1.79-0.08i, 3.88-0.14i)$   & $(-0.81-1.87i, 2.23-1.97i)$   \\ 
    7 & $10.65-0.30i$   & $-6.02-8.88i$     & $(4.71-0.06i, -5.17-0.02i)$   & $(3.60 -7.31i, 5.48+2.18i)$   \\  
    10 & $11.44 -0.31i$   & $-5.05+10.34i$     & $(5.19-0.02i, 4.97+0.01i)$   & $(4.00  +8.08i, -5.43+1.99i)$   \\ 
    \bottomrule
\end{tabular} 
\caption{Selected parameters for interlayer relaxation parameters for the first three shells for TBG at $1.05^\circ$. Terms in the first and second shell are related by $2\pi/3$ rotation symmetry, and they only differ by a phase. Momentum hoppings in the third shell related by $2\pi/3$ rotation (indexed by 7,9,11 or 8,10,12) also only differ by a phase, similar to intralayer terms. }\label{tab:inter_params}
\end{table}

\subsection{Energy scales and comparison with previous models}
The length scale of the TBG moir\'e reciprocal unit cell is approximately  $ \Delta \bvec K := |\bvec K_1 - \bvec K_2 | \approx 0.033 \text{\r{A}}^{-1}$. We can estimate the energy scale of the terms in the Hamiltonian using the length scale of $\Delta \bvec K$. Similar to \cite{Quinn_Kong_Luskin_Watson_2025,Kong2025}, we expand the Hamiltonian by the energy orders of its individual terms. The estimates are reported in \cref{tab:energy}. The chiral and original BM model only contains the first order terms, while the higher-order BM model \cite{Quinn_Kong_Luskin_Watson_2025} contains second order corrections. The structural relaxation effects introduces a new group of intralayer hopping parameters as an additional second order correction, as well as even higher order corrections in the intralayer and interlayer terms \cite{Kang_Vafek_2023,VafekKang2020}.

\begin{landscape}
\begin{table}[h]
\centering
 \renewcommand{\arraystretch}{1.75}
 \begin{tabular}{c  c c c c c c c c c}
 \toprule
 & \multicolumn{3}{c}{First Order} & \multicolumn{3}{c}{Second Order} & \multicolumn{3}{c}{Third Order \& higher} \\
  \cmidrule(lr){2-4} \cmidrule(lr){5-7}  \cmidrule(lr){8-10}
  & Description & Term & $E$ (meV) & Description & Term & $E$ (meV)& Description & Term & $E$ (meV) \\
  \midrule
  \multirow{2}{2cm}{Graphene  dispersion} & \textbf{{Dirac}} & $v \Delta\bvec K$ & 176 & \textbf{\textit{Quad. dispersion AA}} & $v_1 \Delta\bvec K^2$ & 1 \\
  & & & & \textbf{\textit{Quad. dispersion AB}} & $v_2 \Delta\bvec K^2$ & 4 \\
  \midrule
  \multirow{3}{2cm}{Intralayer hops} & & & & Shell 1 AB & $\left| [A_1^{(\ell)}]_{\A\B} \right|$ & 18 & Shell 2 AB  & $\left| [A_7^{(\ell)}]_{\A\B} \right|$ & 0.32 \\
  & & & & Shell 1 $k$-dep & $\left| A_{1,\nabla}^{(\ell)} \right| \Delta \bvec K$ & 1 & Shell 2 $k$-dep  & $\left| A_{7,\nabla}^{(\ell)} \right| \Delta \bvec K$ & 0.02 \\
  & & & & & & & Shell 1 \& 2 AA & $\left| [A_j^{(\ell)}]_{\A\A} \right|$ & 0.02 \\
    \midrule
    \multirow{3}{2cm}{Interlayer hops} & \textbf{Shell 1 AA} & $\left| [\tilde T_1]_{\A\A}\right|$ & 79 & \textbf{\textit{Shell 1 k-dep}}  & $\left| \tilde T_{1,\nabla} \right| \Delta \bvec K$ & 3 & Shell 2 $k$-dep & $\left| \tilde T_{4,\nabla} \right| \Delta \bvec K$ & 0.14\\
    & \textbf{Shell 1 AB}  & $\left| [\tilde T_1]_{\A\B}\right|$ & 113 &
    
    \textbf{\textit{Shell 2}}  & $\left| \tilde T_4\right|$ & 3 & Shell 3 $k$-dep & $\left| \tilde T_{7,\nabla}\right| \Delta \bvec K$ & 0.35 \\
    & & & & Shell 3 & $\left| \tilde T_7\right|$ & 12\\
    \bottomrule
 \end{tabular}
 \caption{Approximate energy scales of the individual terms in the relaxed BM Hamiltonian. The indices correspond to momentum hops in the respective shells (see \cref{fig:mom_hops}.) Terms in bold are included in the original BM model \cite{Bistritzer_MacDonald_2011}. Terms in bold and italic are corrections included in the higher-order BM model \cite{Quinn_Kong_Luskin_Watson_2025}.}\label{tab:energy}
\end{table}
\end{landscape}

\end{document}

%% file: preamble.tex
\usepackage{amsmath,amssymb,amsthm,amsfonts,dsfont}
\usepackage{bm,graphicx,hyperref,cancel,color}
\usepackage[capitalise]{cleveref}
\usepackage{arydshln}
\usepackage{booktabs}
\usepackage{multirow}
\usepackage{braket}
\usepackage{soul}

\usepackage{lipsum}
\usepackage[sort]{cite}
\usepackage[T1]{fontenc}

\newcommand{\bvec}[1]{\ensuremath{\mathbf{#1}}}

\DeclareMathOperator{\tr}{tr}

\renewcommand{\Re}{\operatorname{Re}}
\renewcommand{\Im}{\operatorname{Im}}

\usepackage{etoolbox}
\newtoggle{stdmargins}

\usepackage{tcolorbox}

\usepackage{tikz}
\usetikzlibrary{arrows,shapes,calc,positioning}

\usepackage{xcolor}
\definecolor{scarlet}{RGB}{255, 36, 0}
\definecolor{purp}{RGB}{160, 32, 240}

%% file: paper_preamble.tex
\usepackage{enumitem}

\newtheorem*{assumption*}{Assumptions}

\newtheorem*{theorem*}{Theorem}
\newtheorem*{conjecture*}{Conjecture}
\newtheorem{lemma}{Lemma}[section]
\newtheorem*{lemma*}{Lemma}

\newtheorem*{proposition*}{Proposition}

\newtheorem{remark}[lemma]{Remark}

\theoremstyle{definition}